\documentclass[11pt,a4paper]{article}

\usepackage[english]{babel}

\pdfoutput=1 

\RequirePackage{ifpdf}
\usepackage{amsmath}
\usepackage{mathtools}

\usepackage{axodraw}

\usepackage{jheppub}
\usepackage{pstricks}
\usepackage[final]{pdfpages}
\usepackage{ifpdf}
\usepackage{slashed}
\usepackage{subfiles}

\usepackage{hyperref}

\usepackage{color}
\usepackage{graphics}

\usepackage{etoolbox} 
\usepackage{fixmath}

\usepackage{notoccite}

\usepackage{amsfonts}
\usepackage{float}

\usepackage[normalem]{ulem}

\usepackage[compat=1.1.0]{tikz-feynman}
\tikzfeynmanset{warn luatex=false}
\usetikzlibrary{positioning,arrows}
\usetikzlibrary{decorations.pathmorphing}
\usetikzlibrary{decorations.markings}
\usetikzlibrary{shapes,shadows}

\title{Two-loop mixed QCD-EW corrections to charged current Drell-Yan}

\author[a,b]{Tommaso Armadillo,}
\author[c]{Roberto Bonciani,}
\author[b,d]{Simone Devoto,}
\author[e]{Narayan Rana,}
\author[b,f]{Alessandro Vicini}

\affiliation[a]{Centre for Cosmology, Particle Physics and Phenomenology (CP3), Université catholique de Louvain, Chemin du Cyclotron, 2, B-1348 Louvain-la-Neuve, Belgium}
\affiliation[b]{Dipartimento di Fisica ``Aldo Pontremoli'',
  University of Milano and INFN, Sezione di Milano, I-20133 Milano, Italy}
\affiliation[c]{Dipartimento di Fisica, Universit\`a di Roma ``La Sapienza'' and INFN, Sezione di Roma, I-00185 Roma, Italy}
\affiliation[d]{Department of Physics and Astronomy, Ghent University, 9000 Ghent, Belgium}
\affiliation[e]{School of Physical Sciences, National Institute of Science Education and Research,
An OCC of Homi Bhabha National Institute, 752050 Jatni, India}
\affiliation[f]{CERN, Theoretical Physics Department, CH-1211 Geneva 23, Switzerland}

\emailAdd{tommaso.armadillo@uclouvain.be}
\emailAdd{roberto.bonciani@roma1.infn.it}
\emailAdd{simone.devoto@ugent.be}
\emailAdd{narayan.rana@niser.ac.in}
\emailAdd{alessandro.vicini@mi.infn.it}

\abstract{
  We present the two-loop mixed strong-electroweak virtual corrections
  to the charged current Drell-Yan process.
 The final-state collinear singularities are regularised by the lepton mass.
 The evaluation of all the relevant Feynman integrals,
 including those with up to two different internal massive lines,
 has been worked out relying on semi-analytical techniques,
 using complex-valued masses.
We can provide, at any arbitrary phase-space point, the solution as a power series in the $W$-boson mass, around a reference value.  
Starting from these expansions, we can prepare a numerical grid for any value of the $W$-boson mass within their radius of convergence in a negligible amount of time.
}

\preprint{TIF-UNIMI-2024-3, CERN-TH-2024-047}

\begin{document}
\allowdisplaybreaks[4]
\unitlength1cm
\maketitle
\flushbottom


\def\D{{\cal D}}
\def\DD{\overline{\cal D}}
\def\g{\overline{\cal G}}
\def\gm{\gamma}
\def\M{{\cal M}}
\def\ep{\epsilon}
\def\epm1{\frac{1}{\epsilon}}
\def\epm2{\frac{1}{\epsilon^{2}}}
\def\epm3{\frac{1}{\epsilon^{3}}}
\def\epm4{\frac{1}{\epsilon^{4}}}
\def\unM{\hat{\cal M}}
\def\ashat{\hat{a}_{s}}
\def\asmur{a_{s}^{2}(\mu_{R}^{2})}
\def\sigbar{{{\overline {\sigma}}}\left(a_{s}(\mu_{R}^{2}), L\left(\mu_{R}^{2}, m_{H}^{2}\right)\right)}
\def\sigbarn{{{{\overline \sigma}}_{n}\left(a_{s}(\mu_{R}^{2}) L\left(\mu_{R}^{2}, m_{H}^{2}\right)\right)}}
\def\unas{ \left( \frac{\hat{a}_s}{\mu_0^{\epsilon}} S_{\epsilon} \right) }
\def\rnM{{\cal M}}
\def\bt{\beta}
\def\cD{{\cal D}}
\def\cC{{\cal C}}
\def\ca{\text{\tiny C}_\text{\tiny A}}
\def\cf{\text{\tiny C}_\text{\tiny F}}
\def\ct{{\red []}}
\def\sv{\text{SV}}
\def\murOmu{\left( \frac{\mu_{R}^{2}}{\mu^{2}} \right)}
\def\bb{b{\bar{b}}}
\def\bt0{\beta_{0}}
\def\bt1{\beta_{1}}
\def\bt2{\beta_{2}}
\def\bt3{\beta_{3}}
\def\gm0{\gamma_{0}}
\def\gm1{\gamma_{1}}
\def\gm2{\gamma_{2}}
\def\gm3{\gamma_{3}}
\def\nn{\nonumber}
\def\l{\left}
\def\r{\right}
\def\nn{\nonumber \\&}

\def\asr{\left( \frac{\alpha_s}{4 \pi} \right)}
\def\asrhat{\left( \frac{\hat\alpha_s}{4 \pi} \right)}
\def\aem{\left( \frac{\alpha}{4 \pi} \right)}
\def\smu{\left( \frac{s}{\mu^2} \right)}
\def\J{{\cal J}}
\def\S{{\cal S}}
\def\I{{\cal I}}

\newcommand\as{\alpha_{s}}

\newcommand{\be}{\begin{equation}}
\newcommand{\ee}{\end{equation}}
\newcommand{\bea}{\begin{eqnarray}}
\newcommand{\eea}{\end{eqnarray}}
\newcommand{\smallw}{{\scriptscriptstyle W}}
\newcommand{\mt}{m_t}
\newcommand{\ml}{m_\ell}
\newcommand{\mw}{\mu_\smallw}
\newcommand{\mwsq}{\mu_\smallw^2}
\newcommand{\mwc}{\mu_{\smallw 0}}
\newcommand{\smallz}{{\scriptscriptstyle Z}}
\newcommand{\mz}{\mu_\smallz}
\newcommand{\mzsq}{\mu_\smallz^2}
\newcommand{\mzc}{\mu_{\smallz 0}}
\newcommand{\cmz}{\bar{\mu}_{\smallz}}
\newcommand{\oa}{${\cal O}(\alpha)~$}
\newcommand{\oaa}{${\cal O}(\alpha^2)~$}
\newcommand{\oas}{${\cal O}(\alpha_s)~$}
\newcommand{\oaas}{${\cal O}(\alpha\alpha_s)~$}
\newcommand{\sineffl}{\sin\theta_{eff}^{\ell}\,}
\newcommand{\coseffl}{\cos\theta_{eff}^{\ell}\,}
\newcommand{\seffl}{\sin^2\theta_{eff}^{\ell}\,}
\newcommand{\ceffl}{\cos^2\theta_{eff}^{\ell}\,}
\newcommand{\sw}{s_\smallw\,}
\newcommand{\cw}{c_\smallw\,}
\newcommand{\swd}{s_\smallw^2\,}
\newcommand{\cwd}{c_\smallw^2\,}

\newcommand{\dis}{}
\newcommand{\overbar}[1]{mkern-1.5mu\overline{\mkern-1.5mu#1\mkern-1.5mu}\mkern
1.5mu}

\tikzset{
particle/.style={thick,draw=blue, postaction={decorate},
    decoration={markings,mark=at position .5 with {\arrow[blue]{triangle 45}}}},
uup/.style={draw=blue, postaction={decorate},
    decoration={markings,mark=at position .5 with {\arrow[blue]{stealth}}}},
top/.style={draw=red, postaction={decorate},
    decoration={markings,mark=at position .5 with {\arrow[red]{stealth}}}},
photon/.style={decorate, draw=black,
    decoration={coil,aspect=0,segment length=5pt,amplitude=1.5pt}},
Zboson/.style={decorate, draw=red,
    decoration={coil,aspect=0,segment length=5pt,amplitude=1.5pt}},
Wboson/.style={decorate, draw=mygreen,
    decoration={coil,aspect=0,segment length=5pt,amplitude=1.5pt}},
gluon/.style={decorate, draw=black,
    decoration={coil,aspect=0.5,segment length=2pt,amplitude=2pt}},
fermion/.style={draw=blue,
      postaction={decorate},decoration={markings,mark=at position .55
        with {\arrow[draw=blue]{>}}}},
hadron/.style={draw=black,
      postaction={decorate},decoration={markings,mark=at position .55
        with {\arrow[draw=black]{>}}}},
lepton/.style={draw=violet,
      postaction={decorate},decoration={markings,mark=at position .55
        with {\arrow[draw=violet]{>}}}},
neutrino/.style={draw=violet,
      postaction={decorate},decoration={markings,mark=at position .55
        with {\arrow[draw=violet]{>}}}},
vector/.style={decorate, decoration={snake}, draw},
crayon2b/.style={draw=blue!40!white, line width=2pt, line join=round,
  decoration={random steps, segment length=0.25pt, amplitude=0.5pt}, decorate},
crayon2g/.style={draw=green!40!white, line width=2pt, line join=round,
  decoration={random steps, segment length=0.25pt, amplitude=0.5pt}, decorate},
crayon2r/.style={draw=red!40!white, line width=2pt, line join=round,
  decoration={random steps, segment length=0.25pt, amplitude=0.5pt}, decorate},
crayon5/.style={line width=5pt, line join=round,
  decoration={random steps, segment length=0.5pt, amplitude=1.5pt}, decorate}
 }

\section{Introduction}

\setcounter{equation}{0}
\label{sec:intro}

The hadro-production of lepton-antilepton or lepton-antineutrino pairs, known, respectively, as neutral current (NC) and charged current (CC) Drell-Yan (DY) processes \cite{Drell:1970wh}, are crucial at hadron colliders since they provide the environment for a precise study of the gauge sector of the Standard Model (SM) of fundamental interactions. In particular, the NC DY is important for the determination of the $Z$-boson mass and the effective weak mixing angle, and the CC DY is important for the determination of the $W$-boson mass. These three electroweak (EW) parameters are known at the moment with small relative errors, 0.01\% for the boson masses and 0.1\% for the effective weak mixing angle \cite{Group:2012gb,Aaboud:2017svj,Aaltonen:2018dxj,ATLAS:2018gqq,CMS-PAS-SMP-22-010}, originating from a fitting procedure between measured kinematical distributions and generated theoretical templates \cite{CarloniCalame:2016ouw,Bagnaschi:2019mzi,Behring:2021adr,Rottoli:2023xdc,Torrielli:2023tiz}.

Due to highly precise experimental data, the control on the precision of these parameters forces the theoretical predictions to involve higher-order perturbative corrections.
Next-to-leading-order~(NLO)~\cite{Altarelli:1979ub} and next-to-next-to-leading-order~(NNLO)~\cite{Hamberg:1990np,Harlander:2002wh} QCD corrections to the total production cross section of a $Z$ or $W$ boson have been computed more than 30 years ago. The inclusion of the following order in the strong coupling constant $\as$, the next-to-next-to-next-to-leading-order (N$^3$LO), has been recently completed for the inclusive production of a virtual photon~\cite{Duhr:2020seh}, a $W$~\cite{Duhr:2020sdp}, and a $Z$ boson~\cite{Duhr:2021vwj}. The NNLO QCD corrections have been implemented at the fully differential level in~\cite{Anastasiou:2003yy,Anastasiou:2003ds,Melnikov:2006kv,Catani:2009sm,Catani:2010en}.
Fiducial cross sections for the NC DY and CC DY processes at N$^3$LO QCD have been computed in~\cite{Camarda:2021ict,Chen:2022cgv,Neumann:2022lft,Campbell:2023lcy}. Finally, more exclusive observables, as the di-lepton rapidity distribution and the transverse mass and charge asymmetry, have been computed to third order in QCD in \cite{Chen:2021vtu} and \cite{Chen:2022lwc}, respectively.
The resummation of logharithmically enhanced terms near the production threshold has been considered in  \cite{Moch:2005ky,Laenen:2005uz,Ravindran:2005vv,Ravindran:2006cg,deFlorian:2012za,Ahmed:2014cla,Catani:2014uta,Li:2014afw,Ajjath:2020ulr}.
NLO EW corrections are known both for $W$~\cite{Dittmaier:2001ay,Baur:2004ig,Zykunov:2006yb,Arbuzov:2005dd,CarloniCalame:2006zq}, and for $Z$ production~\cite{Baur:2001ze,Zykunov:2005tc,CarloniCalame:2007cd,Arbuzov:2007db,Dittmaier:2009cr}.

The combination of NLO QCD and EW results with QCD and QED Parton Showers, has been presented in~\cite{Bernaciak:2012hj,Barze:2012tt,Barze:2013fru,Frederix:2018nkq,Chiesa:2024qzd}. 
At the required level of accuracy, NNLO mixed QCD$\times$EW corrections are also relevant in particular for kinematical distributions \cite{CarloniCalame:2016ouw}. 

For on-shell inclusive $Z$ production, the NNLO QCD$\times$QED corrections have been calculated in ~\cite{deFlorian:2018wcj}.
Exclusive on-shell $Z$-boson production has been considered in ~\cite{Delto:2019ewv} and in~\cite{Cieri:2020ikq}.
The complete NNLO mixed QCD$\times$EW corrections to the total cross section of production of an on-shell $Z$ boson have been presented in~\cite{Bonciani:2016wya,Bonciani:2019nuy,Bonciani:2020tvf,Bonciani:2021iis}, while in~\cite{Buccioni:2020cfi} and ~\cite{Behring:2020cqi} the same corrections have been evaluated differentially in the final-state leptonic variables for on-shell $Z$ and $W$ production respectively.

Mixed QCD$\times$EW corrections at the $W$ or $Z$ resonances has been evaluated in pole approximation in~\cite{Dittmaier:2014qza,Dittmaier:2015rxo}. In~\cite{Dittmaier:2020vra} the gauge-invariant set of corrections that include a closed fermionic loop, at ${\cal O}(n_F\as \alpha)$, with $\alpha$ the fine-structure constant and $n_F$ the number of active flavors, were evaluated for off-shell $Z$ and $W$ production.
In the case of NC DY, recently, differential distributions and the forward-backward asymmetry have been evaluated in pole approximation in~\cite{Dittmaier:2024row}. 

The complete set of NNLO QCD$\times$EW corrections to the NC DY process $pp\to \ell^+\ell^-+X$ is available and has been presented in~\cite{Bonciani:2021zzf}, for massive leptons, and in~\cite{Buccioni:2022kgy} for massless leptons.
The calculation in~\cite{Bonciani:2021zzf} is based on the exact two-loop amplitudes presented in~\cite{Armadillo:2022bgm}, evaluated through the reduction to the master integrals~\cite{Bonciani:2016ypc} and a semi-analytical implementation of the differential equations method for their calculation~\cite{Moriello:2019yhu,Hidding:2020ytt,Armadillo:2022ugh}. The IR singularities have been treated in the $q_{\rm T}$ subtraction formalism \cite{Buonocore:2021tke, Camarda:2021jsw}, implemented in the MATRIX framework \cite{Grazzini:2017mhc}. 
The calculation in~\cite{Buccioni:2022kgy} is based on the exact amplitudes presented in~\cite{Heller:2020owb} expressed in terms of master integrals evaluated in analytic polylogarithmic form~\cite{Heller:2019gkq,Hasan:2020vwn}.

For the CC DY process $pp\to \ell \nu_\ell+X$, the mixed QCD$\times$EW corrections have been computed in~\cite{Buonocore:2021rxx}, in the case of massive lepton, with approximated two-loop virtual corrections. While the full set of double-real and real-virtual contributions due to initial and final state radiation have been included in the analysis in exact form, the finite part of the two-loop virtual amplitudes has been evaluated in pole approximation.
Large mixed QCD$\times$EW effects might develop in the large lepton-pair transverse/invariant mass limit, where the pole approximation is expected to become less accurate, and are relevant for new physics searches.
In this paper, we calculate the virtual two-loop amplitude at ${\cal O}(\as \alpha)$ exactly. Final-state collinear divergences are regularized by the finite mass of the final-state lepton. We keep the exact dependence on the masses of the vector bosons and we find expressions that are valid in the whole physical phase-space.
In Section \ref{sec:frame},
we provide the framework of our computation, with the description of the ultraviolet (UV)
renormalisation procedure.
In Section \ref{sec:ampl},
the details of the computation are presented,
from the amplitude level up to the renormalised matrix elements.
In Section \ref{sec:infra},
the infrared (IR) subtraction procedure is discussed and we describe how we obtain the UV renormalised IR subtracted finite remainder.
In Section \ref{sec:results} we present our numerical results.
Finally, in Section \ref{sec:conclusions}
we draw our conclusions.
The unrenormalised matrix elements in terms of master integrals
are provided as ancillary material.
Appendix \ref{app:readme} gives the reader the information needed to make use of the files.

\section{Framework of the Calculation}

\label{sec:frame}
\subsection{The process}
The process under study is the production of a lepton-neutrino pair in quark-antiquark annihilation
\begin{equation}
 u(p_1) + \bar{d}(p_2) \rightarrow \nu_\ell(p_3) + \ell^{+} (p_4) \,.
\label{eq:process}
\end{equation}
We consider the two-loop mixed QCD-EW corrections to this reaction, with explicit results for massless initial-state quarks.
This partonic process has a 
bare amplitude\footnote{We  denote by \textit{hat} the bare quantities.}, which
admits a perturbative expansion in the two coupling constants
\begin{align}
 |\unM \rangle = |\unM^{(0)} \rangle + \asrhat |\unM^{(1,0)} \rangle + \hat\aem |\unM^{(0,1)} \rangle + \asrhat \hat\aem |\unM^{(1,1)} \rangle + \cdots
\end{align}
We compute the interference terms:
\begin{equation}
  \langle \unM^{(0)} | \unM^{(1,0)} \rangle \,, ~~
  \langle \unM^{(0)} | \unM^{(0,1)} \rangle \,, ~~
  \langle \unM^{(0)} | \unM^{(1,1)} \rangle \,,
  \label{eq:interferences}
\end{equation}
which contribute  to the unpolarised squared matrix element
of the process in Eq.~\eqref{eq:process}, 
at \oaas\!\!.
The virtual corrections to any scattering process amplitude are in general affected by singularities of UV and IR type, which we regularise working in $d=4-2\varepsilon$ space-time dimensions.
We apply the
integration-by-parts (IBP) \cite{Tkachov:1981wb,Chetyrkin:1981qh} and
Lorentz invariance (LI) \cite{Gehrmann:1999as} identities
to reduce all the scalar Feynman integrals appearing in the interference terms (\ref{eq:interferences}) to a limited set, the so-called Master Integrals (MIs).
The final expression can then be cast as the sum of the MIs multiplied by their associated coefficients.
The latter depend, as rational functions, on the kinematical invariants of the process, the masses of the internal particles exchanged in the Feynman diagrams, and the dimensional regularization parameter $\varepsilon$.

The $d$-dimensional nature of an EW computation performed in dimensional regularisation requires the
extension to $d$ dimensions of the inherently 4-dimensional object $\gamma_5$.
In this work we follow the same strategy outlined in~\cite{Armadillo:2022bgm}, preserving the anti-commutation proprieties of $\gamma_5$ in a fixed-point prescription. We refer the reader to~\cite{Armadillo:2022bgm} for more details on the procedure.

The cancellation of the UV divergences follows according to
a standard renormalisation procedure. The resulting amplitudes are expressed in terms of renormalised parameters,
in turn related to measurable quantities.
The IR divergences have a universal structure, arising from the behavior of the amplitudes in the soft and/or collinear limit.
The universality of the divergent factors
has been discussed at length in the literature and
allows us to prepare a subtraction term, which is independent of the details of the full two-loop calculation, and depends only on the lower order amplitudes.
We exploit the universality property to check the consistency of our computation,
for the prescription related to the Dirac $\gamma_5$ matrix.
We write the UV-renormalised amplitude as a Laurent expansion in $\varepsilon$.
We have, at one- and two-loop level respectively:
\bea
  \langle {\cal M}^{(0)} | {\cal M}^{(0,1)} \rangle \, &=& \sum_{i=-2}^{2} \varepsilon^i C^{(0,1)}_i(s,t,\mw,\mz,\ml)\, ,\label{i01}\\
  \langle {\cal M}^{(0)} | {\cal M}^{(1,0)} \rangle \, &=& \sum_{i=-2}^{2} \varepsilon^i C^{(1,0)}_i(s,t,\mw,\mz,\ml)\, ,\label{i10}
\eea
and
\bea
  \langle {\cal M}^{(0)} | {\cal M}^{(1,1)} \rangle \, &=& \sum_{i=-4}^{0} \varepsilon^i C^{(1,1)}_i(s,t,\mw,\mz,\ml)\,\label{i11} .
\eea
In Eq.~(\ref{i11}) only terms up to $\varepsilon^0$ have been kept,
because we are interested in the finite contributions at NNLO only.
The higher powers of $\varepsilon$ would be relevant for higher-order calculations.
The lepton mass is labelled by $m_l$.
Given the mass and decay width  $M_V,\Gamma_V$ of the gauge boson $V$ ($V=W,Z$),
we define $\mu_V$ 
as the position in the complex plane of the pole  of the boson propagator
\be
\mu_V^2=M_V^2-i M_V\Gamma_V\;.
\ee
The Mandelstam variables are defined as:
\begin{equation}
 s = (p_1+p_2)^2, \, t = (p_1-p_3)^2, \, u = (p_2-p_3)^2 \,\, {\rm with} \,\, s+t+u= m_l^2 \,,
\end{equation}
while the on-shell conditions of the external particles are:
\begin{equation}
 p_1^2 = p_2^2 = p_3^2 = 0; ~ p_4^2  = m_l^2;
\end{equation}
The main result of this paper is
the IR-subtracted expressions that can be obtained by the combination of
the two-loop coefficients $C_i^{(1,1)}$ in Eq.~(\ref{i11}) with the universal subtraction term. The latter is based on the availability of the Born and the one-loop matrix elements in Eqs.~(\ref{i01},\ref{i10}).

\subsection{Ultraviolet renormalisation}
\label{sec:UV}
The NC DY process renormalisation, at \oaas\!\!, has already been discussed in detail in~\cite{Dittmaier:2020vra, Armadillo:2022bgm}.
For the CC DY process we follow similar steps, that we summarize in this section.

\subsubsection{Charge renormalisation}
The $SU(2)_L$ and $U(1)_Y$ bare gauge couplings $g_0, g'_0$ and the Higgs doublet vacuum expectation value $v_0$ can be related, with appropriate counterterms, to their renormalised counterparts $g,g',v$.
We link $g,g',v$ to a set of three measurable quantities:
for instance $G_\mu,\mw,\mz$ (with $G_\mu$ the Fermi constant)
or $e,\mw,\mz$ (with $\alpha=e^2/(4\pi)$ the fine structure constant and $e$ the positron charge).
The choice of the three measurable quantities defines the so called input scheme.
The relation between bare and renormalised input parameters is
\bea
\mwc^2&=&\mwsq+\delta\mwsq,\quad  \mzc^2=\mzsq+\delta\mzsq,\quad e_0=e+\delta e\, ,
\label{eq:dmass}
\eea
The on-shell electric charge counterterm $\delta e$ at \oaa and \oaas has been discussed in~\cite{Degrassi:2003rw},
from the study of the Thomson scattering.
The mass counterterms $\delta\mwsq,\, \delta\mzsq$ in the complex mass scheme  \cite{Denner:2005fg}
have been presented in~\cite{Dittmaier:2020vra}.
In terms of the transverse part of the unrenormalised $VV$ gauge boson self-energies, they are defined as follows:
\be
\delta\mu_V^2 = \Sigma_{T}^{VV}(\mu_V^2)\, ,
\ee
at the pole in the complex plane $q^2=\mu_V^2$ of the gauge boson propagator, with $V=W,Z$.
From the study of the muon-decay amplitude, we derive the following relation
\be
\frac{G_\mu}{\sqrt{2}}=
\frac{\pi \alpha}{2}\frac{\mzsq}{ \mwsq (\mzsq-\mwsq)}\left( 1+ \Delta r \right)\, .
\ee
The finite correction $\Delta r$ has been introduced,
with real gauge boson masses, in~\cite{Sirlin:1980nh}
and its \oaas corrections were presented in~\cite{Kniehl:1989yc,Djouadi:1993ss}. We evaluate it here with complex-valued masses.

We consider now the bare couplings which appear at tree-level in the interaction of the photon and $W$ boson with fermions.
The UV divergent correction factors $\delta g_W^{G_\mu}$ and $\delta g_W^{\alpha}$ contribute to the
charge renormalisation of the $Wf\bar f'$ vertex in the  $(G_\mu,\mw,\mz)$ and $(\alpha,\mw,\mz)$ input schemes respectively as
\bea
g_0 &=&
\sqrt{4\sqrt{2} G_\mu \mw^2}
\left[ 1
- \frac12 \Delta r
+ \frac{\delta e}{e} 
+ \frac12 \frac{\mw^2}{\mz^2-\mw^2} 
\left(\frac{\delta \mw^2}{\mw^2}
-\frac{\delta \mz^2}{\mz^2}
\right)\right]  \\
& \equiv &
\sqrt{4\sqrt{2} G_\mu \mw^2} \left(
1 + \delta g_W^{G_\mu} \right)~~~~~~~~~
\label{eq:shiftsWGmu}
\eea
and
\bea
g_0 &=&
e \sqrt{\frac{\mz^2}{\mz^2-\mw^2}}
\left[ 1
+ \frac{\delta e}{e} 
+ \frac12 \frac{\mw^2}{\mz^2-\mw^2} 
\left(\frac{\delta \mw^2}{\mw^2}
-\frac{\delta \mz^2}{\mz^2}
\right)\right] \\
& \equiv &
e \sqrt{\frac{\mz^2}{\mz^2-\mw^2}}
\left(
1 + \delta g_W^{G_\alpha} \right) \, .
\label{eq:shiftsWalpha}
\eea
Working out the explicit expression of $\delta g_Z^{G_\mu}$, the dependence on the electric charge counterterm cancels out.

The counterterm contributions to the renormalised amplitude are obtained by replacing
the bare couplings in the lower order amplitudes with the expressions
presented in Eqs.~(\ref{eq:shiftsWGmu}-\ref{eq:shiftsWalpha})
and expanding $\delta g_W$ up to the relevant perturbative order.
We show in the next Section how  $\delta g_W$ enters in the renormalisation of the $W$ boson propagators.

\subsubsection{Renormalisation of the gauge boson propagators}
\label{sec:gaugeren}
The renormalised gauge boson self-energies are obtained, at \oa\hspace{-0.5em},
by combining the unrenormalised self-energy expressions with the mass and wave function counterterms.
In the full calculation, we never introduce wave function counterterms on the internal lines,
because they would systematically cancel
against a corresponding factor stemming from the definition of the renormalised vertices.
We exploit instead the relation in the SM
between the wave function and charge counterterms \cite{Denner:2019vbn}
and we directly use the latter to define the renormalised self-energies.
We obtain:
\bea
\Sigma_{R,T}^{WW}(q^2) &=&
\Sigma_T^{WW}(q^2) -\delta\mu_W^2 + 2\, (q^2-\mwsq)\,\delta g_W \, ,
\eea
where $\Sigma_T^{VV}$ and $\Sigma_{R,T}^{VV}$ are the transverse part of the bare and renormalised $VV$ vector boson self-energy.
The charge counterterms have been defined in Eqs.~(\ref{eq:shiftsWGmu}-\ref{eq:shiftsWalpha}).
At \oaas the structure of these contributions does not change:
the corrections to the gauge boson self-energies
stem from a quark loop with one internal gluon exchange and,
in addition, from the \oas mass renormalisation of the quark lines in the one-loop self-energies.

The expression of the two-loop Feynman integrals required for the evaluation of the
\oaas correction to the gauge boson propagators and all the needed counterterms
can be found in~\cite{Kniehl:1989yc,Djouadi:1993ss,Dittmaier:2020vra}.
\section{The UV-renormalised unsubtracted amplitude}

\label{sec:ampl}
\subsection{Generation of the full amplitude and evaluation of the interference terms}

\begin{figure}
\centering
\includegraphics[scale=0.4]{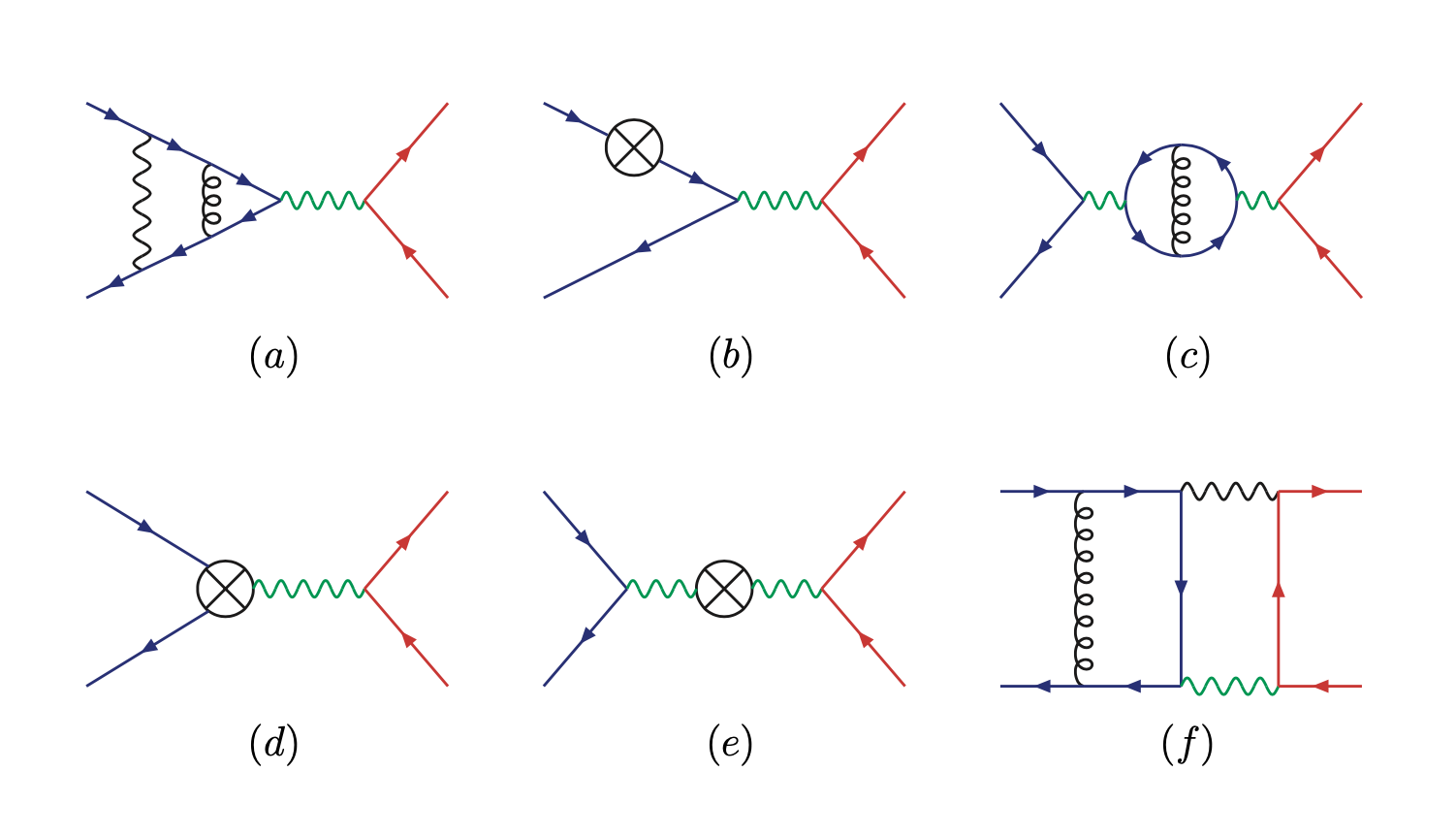}
\caption{Sample Feynman diagrams of two-loop corrections and associated two-loop counterterms.
\label{fig:diagsamples}}
\end{figure}

\begin{figure}
\centering
\includegraphics[scale=0.4]{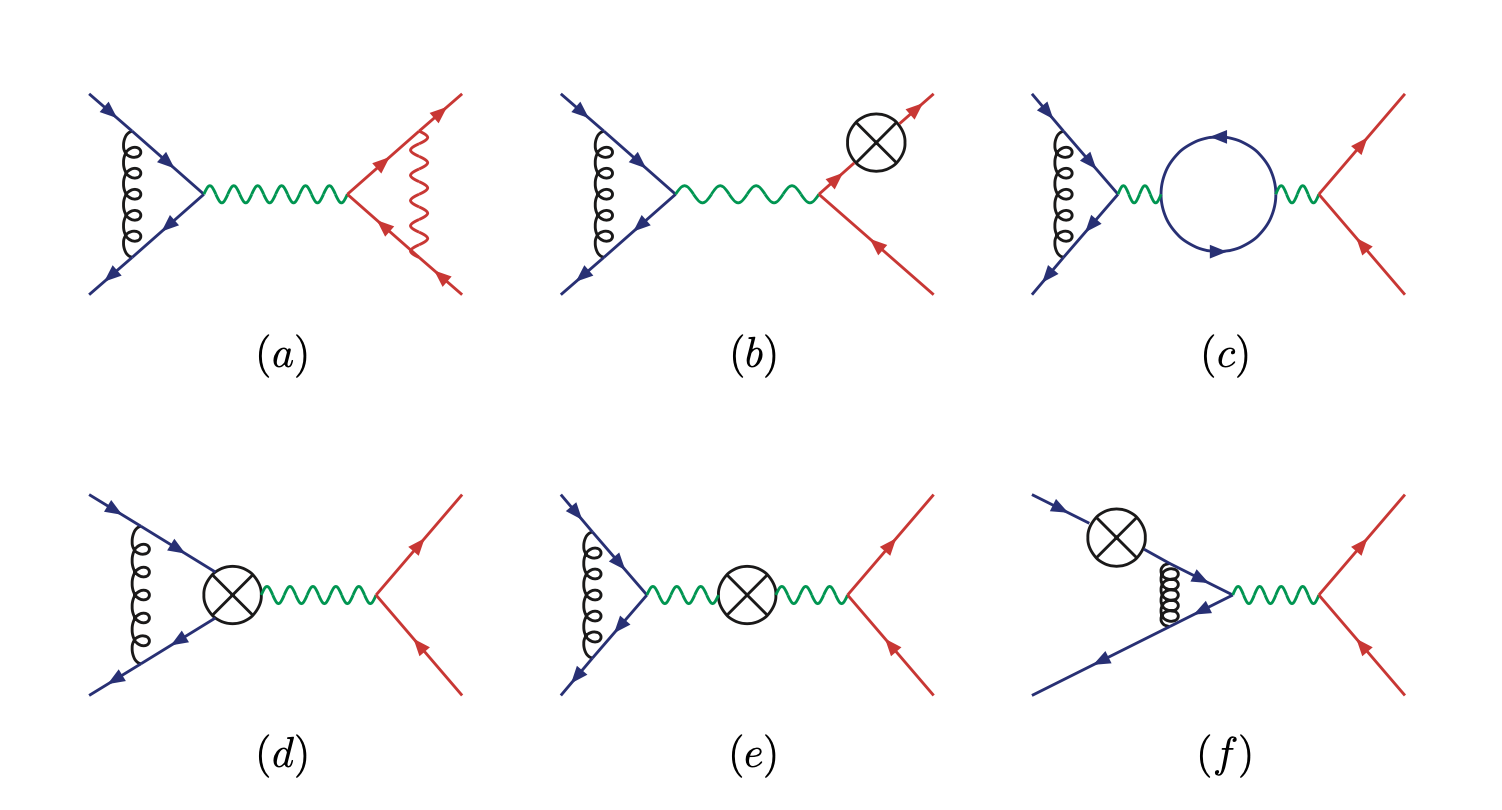}
\caption{Sample Feynman diagrams of factorisable corrections, including one-loop counterterm corrections.
\label{fig:diagsamplesCT}}
\end{figure}

Different kinds of Feynman diagrams contribute,
at \oaas\!\!, to the scattering amplitude.
We present\footnote{
The straight line with an arrow represents a fermion line
(blue: quark, red: leptons),
the wavy lines represent EW gauge bosons (green: $W$, red: $Z$, black: photon)
and the spiral coils represent gluons.
}
in Figures \ref{fig:diagsamples} and \ref{fig:diagsamplesCT}
a few representative examples.
The initial state
two-loop vertex corrections (Figure~\ref{fig:diagsamples}-(a)),
are combined with
the two-loop external quark wave function corrections
(Figure~\ref{fig:diagsamples}-(b))
and with
the one-loop external quark wave function correction
with initial-state QCD vertex correction (Figure~\ref{fig:diagsamplesCT}-(f)).
The combination
yields an UV-finite, but still IR-divergent, result.
We treat as a separate subset the sum of 
two-loop $W$ self-energy,
together with the corresponding two-loop mass counterterms
and with the insertion of charge renormalisation constants in the initial and final state vertices.
Representative diagrams are shown in Figures \ref{fig:diagsamples}-(c),(d),(e).
The combination yields an UV- and IR-finite contribution.
An example of a two-loop box with the exchange of one $W$ and one neutral EW boson
is given in Figure \ref{fig:diagsamples}-(f).

The factorisable contributions 
are schematically represented in
Figures~\ref{fig:diagsamplesCT}-(a),(f).
They all include, at \oaas\!\!, 
an initial-state QCD vertex correction.
The second factor can be, alternatively:
the final-state EW vertex corrections,
the external lepton wave function correction,
the one-loop $W$ self-energy corrections,
the one-loop $W$ mass renormalisation counterterms,
the one-loop external quark wave function correction and the one-loop charge renormalisation contributions.
The properties of the background field gauge (BFG) \cite{Denner:1994xt} allow to identify UV-finite  Feynman diagrams combinations.

We follow a general procedure to compute the bare matrix elements up to two-loop order.
We generate the Feynman diagrams
with two completely independent approaches,
one based on 
{\sc FeynArts} \cite{Hahn:2000kx}
and a second one on 
{\sc QGRAF}~\cite{Nogueira:1991ex}.
The in-house {\sc Mathematica} package {\sc ABISS} has been used
to perform the Lorentz and Dirac algebra and obtain the interference between the virtual corrections and the Born amplitude. 
It has been compared with the independent results of a second set of procedures
written in {\sc FORM} \cite{Vermaseren:2000nd}. 
The large number of scalar Feynman integrals appearing in the intermediate expressions
is reduced to a smaller set of MIs by means of IBP and LI identities.
The reduction algorithms are implemented in two independent in-house programs based on
{\sc Kira} \cite{Maierhofer:2017gsa} and {\sc LiteRed}~\cite{Lee:2013mka, Lee:2012cn}.
We have cross-checked at one- and two-loop level
the corresponding expressions,
finding perfect agreement.
Additional technical details, common to the NC and CC DY processes, are described in~\cite{Armadillo:2022bgm}.

We retain the exact dependence on the lepton mass $\ml$, both at the one- and two-loop level, only in the diagrams where an internal photon couples to the final-state lepton, featuring the complete set of contributions enhanced by $\log(\ml^2)$. The lepton mass regulates in this case the final-state collinear divergences.
These diagrams also yield  a gauge-invariant subset of terms proportional to $\ml^2/\mw^2$, which are however phenomenologically negligible. In all the other cases, we do not have final-state IR divergences, and therefore we can keep the lepton massless. The use of this approximation reduces by one the number of independent mass scales in the computation and thus its complexity, 
allowing for a faster evaluation of the amplitude.

\subsection{Integral families and the representation of the result in terms of MIs}
\label{sec:masterintegrals}

The scattering amplitude, after the computation of the interference terms, appears as a sum of scalar Feynman integrals. 

In order to classify all the integrals entering in our computation, we need 11 different integral families.
Among them, 7 can be defined exactly as the ones introduced in~\cite{Armadillo:2022bgm} for the case of the NC DY process:
\begin{align}
  \label{eq:Basis_ncdy}
  {\rm B}_{0} &: \{ \cD_1, \cD_2, \cD_{12}, \cD_{1;1}, \cD_{2;1}, \cD_{1;12}, \cD_{2;12}, \cD_{1;3}, \cD_{2;3} \}\nonumber\\
  {\rm B}_{1} &: \{ \cD_1, \cD_2, \cD_{12}, \cD_{1;1}, \cD_{2;1}, \cD_{1;12}, \cD_{12;2}, \cD_{1;3}, \cD_{2;3} \}\nonumber\\
  {\rm B}_{11} &: \{ \cD_1, \cD_2, \cD_{12}-\mu_V^2, \cD_{1;1}, \cD_{2;1}, \cD_{1;12}, \cD_{2;12}, \cD_{1;3}, \cD_{2;3} \}
              \nonumber\\
  {\rm B}_{12} &: \{ \cD_1, \cD_2, \cD_{12}, \cD_{1;1}-\mu_V^2, \cD_{2;1}, \cD_{1;12}, \cD_{2;12}, \cD_{1;3}, \cD_{2;3} \}
              \nonumber\\
  {\rm B}_{13} &: \{ \cD_1, \cD_2, \cD_{12}-\mu_V^2, \cD_{1;1}, \cD_{2;1}, \cD_{1;12}, \cD_{12;2}, \cD_{1;3}, \cD_{2;3} \}
              \nonumber\\
{\rm B}_{14} &: \{ \cD_1, \cD_2-\mu_V^2, \cD_{12}, \cD_{1;1}, \cD_{2;1}, \cD_{1;12}, \cD_{2;12}, \cD_{1;3}, \cD_{2;3} \}
              \nonumber\\
 {\rm B}_{18} &: \{ \cD_1, \cD_2, \cD_{12}, \cD_{1;1}, \cD_{2;1}, \cD_{1;12}, \cD_{2;12}, \cD_{1;3}-\mu_V^2, \cD_{2;3}-\mu_V^2 \}\,,              
\end{align}
where $V$ can be either $Z$ or $W$, and we have defined the denominators $\cD$ as follows:
\begin{equation}
  \cD_{i} = k_{i}^2, \cD_{ij} = (k_i-k_j)^2,
  \cD_{i;j} = (k_i-p_j)^2, \cD_{i;jl} = (k_i-p_j-p_l)^2, \cD_{ij;l} = (k_i-k_j-p_l)^2
  \label{eq:denominators}
\end{equation}
The remaining 4 integral families only appear for the case of CC DY:
\begin{align}
  \label{eq:Basis_ccdy}
  \tilde{\rm B}_{14} &: \{ \cD_1, \cD_2-\mu_V^2, \cD_{12}, \cD_{1;1}, \cD_{2;1}, \cD_{1;12}, \cD_{2;12}, \cD_{1;3}, \cD_{2;3}-m_\ell^2 \}
              \nonumber\\
  \tilde{\rm B}_{14p} &: \{ \cD_1, \cD_2-\mu_V^2, \cD_{12}, \cD_{1;2}, \cD_{2;2}, \cD_{1;12}, \cD_{2;12}, \cD_{1;3}, \cD_{2;3}-m_\ell^2 \}
              \nonumber\\
 \tilde{\rm B}_{16} &: \{ \cD_1, \cD_2-\mu_{V_1}^2, \cD_{12}, \cD_{1;1}, \cD_{2;1}, \cD_{1;12}, \cD_{2;12}-\mu_{V_2}^2, \cD_{1;3}, \cD_{2;3} \}
              \nonumber\\
  \tilde{\rm B}_{16p} &: \{ \cD_1, \cD_2-\mu_{V_1}^2, \cD_{12}, \cD_{1;2}, \cD_{2;2}, \cD_{1;12}, \cD_{2;12}-\mu_{V_2}^2, \cD_{1;3}, \cD_{2;3} \}\,,  
\end{align}
where $V_1$ and $V_2$ are two different vector bosons.

\begin{figure}
    \centering
    \includegraphics[width=1.0\textwidth]{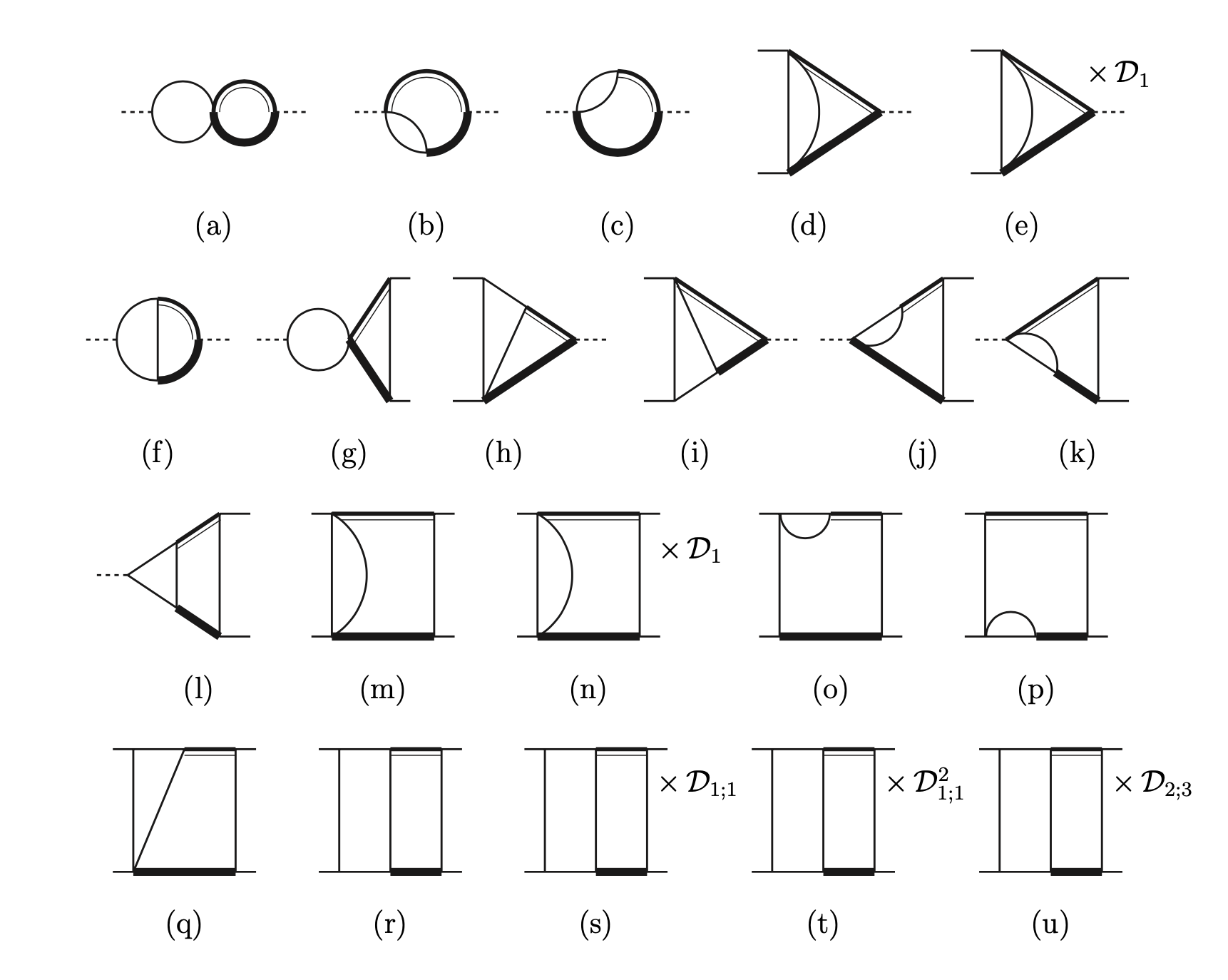}
    \caption{Master Integrals with two different internal masses, belonging to topology $\tilde{B}_{16}$. Thin plane lines represent massless particles. Thick and double lines represent massive particles. External dashed lines indicate the Mandelstam invariant $s$.}
    \label{fig:6666}
\end{figure}

\begin{figure}
    \centering
    \includegraphics[width=1.0\textwidth]{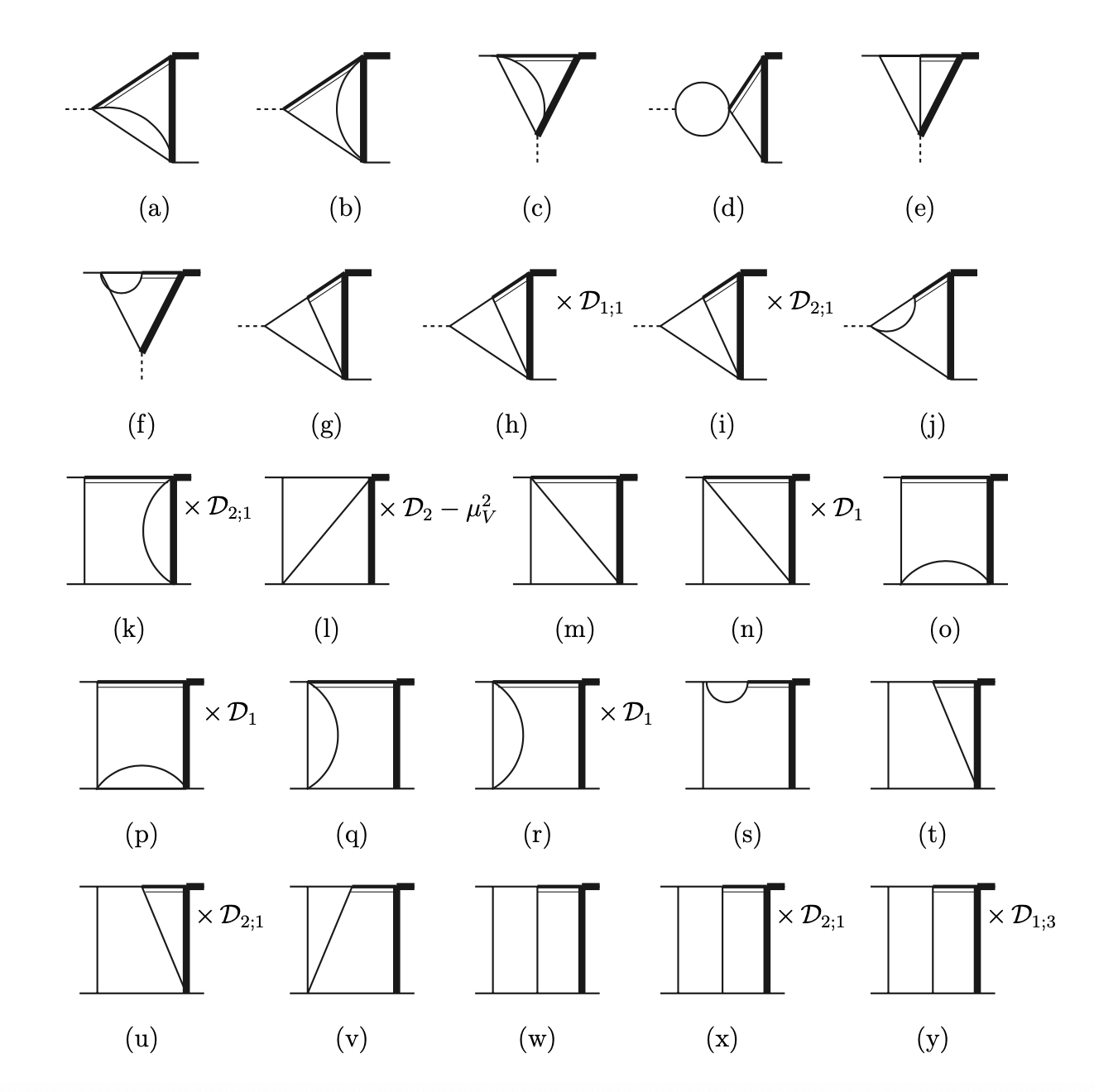}
    \caption{Master Integrals with two different  masses, belonging to topology $\tilde{B}_{14}$. Thin plane lines represent massless particles. Thick and double lines represent massive particles. External dashed lines indicate the Mandelstam invariant $s$.}
    \label{fig:7777}
\end{figure}

The integral families listed in Eq.~(\ref{eq:Basis_ccdy}), despite being specific to the charged current case, can also be seen as an extension of integral families already appearing in the neutral current process, where the value of the masses in the propagators has been modified.
In particular, $\tilde{\rm B}_{16}$ describes box integrals with a $W$-$Z$ exchange and is the natural extension of ${\rm B}_{16}$, already introduced in~\cite{Armadillo:2022bgm} to describe the $Z$-$Z$ and $W$-$W$ box integrals.
The latter can in fact be obtained from $\tilde{\rm B}_{16}$ in the equal mass limit $\mu_{V_1}\to\mu_V$, $\mu_{V_2}\to\mu_V$.
Similarly, by considering a massless lepton, $m_\ell=0$, in the integral family $\tilde{\rm B}_{14}$, we retrieve the integral family ${\rm B}_{14}$ listed in Eq.~(\ref{eq:Basis_ncdy}).

In the NC DY case, we introduced the integral family ${\rm B}_{14}$ instead of the more general $\tilde{\rm B}_{14}$ because the final state collinear singularities, like e.g. those stemming from the $\gamma$-$Z$ boxes, were cancelling between different contributions~\cite{Frenkel:1976bj}. 
We were thus able to take the massless limit for this subset of diagrams.
The integral family ${\rm B}_{14}$ yields collinear poles in dimensional regularization, while such singularities appear in $\tilde{\rm B}_{14}$ as 
lepton-mass logarithms.

After performing the reduction to MIs, the bare result can be cast in the following, compact form:
\begin{equation}
\langle \unM^{(0)} | \unM^{(1,1)} \rangle \, = \sum_{k=1}^{274} c_k(s,t,\{m_i\},\varepsilon) I_k(s,t,\{m_i\},\varepsilon)
\label{eq:sumint}
\end{equation}
where $c_k$ are rational coefficients, functions of the kinematical invariants ($s,t$),
the set of real and complex masses $\{m_i\}$
and of the regularisation parameter $\varepsilon$, while $I_k$ are the MIs\footnote{This intermediate result is provided as an ancillary file attached to this paper.}.
In addition to the MIs employed for the NC DY, in the case of the CC DY we have to evaluate two more sets. In Figure \ref{fig:6666} we show the additional MIs belonging to topology $\tilde{B}_{16}$, that, therefore, have two different masses in the propagators. In Figure \ref{fig:7777}, instead, we show the MIs of topology $\tilde{B}_{14}$, that have two different masses in the propagators and a massive external leg.

\subsection{Numerical evaluation of MIs via series expansions}
\label{sec:semianalytical}

The final step of the calculation consists in the numerical evaluation of the MIs.
The additional complication of the charged-current calculation, with respect to the neutral-current one presented in~\cite{Armadillo:2022bgm}, is given by the presence of the $\tilde{B}_{14}$ and $\tilde{B}_{16}$ topologies, that are two-loop boxes with two internal (and one external in the case of $\tilde{B}_{14}$), and different, massive lines. In the case of $\tilde{B}_{16}$,
the analytical result for the equal masses case has been presented in~\cite{Bonciani:2016ypc} in terms of Chen-iterated integrals and an analytical expression in terms of GPLs has been found in~\cite{Heller:2019gkq}. 
However, for the different masses case, no analytical expressions are available in the literature. We have opted to compute the integrals using the series expansion approach. 

We have generated the differential equations using in-house {\sc Mathematica} routines, which rely on the external package {\sc LiteRed} for computing the derivatives with respect to kinematic invariants $s$ and $t$ and on the IBPs relations provided by {\sc Kira}.
The boundary conditions for all the MIs have been computed in a point in the physical region by using {\sc AMFlow} \cite{Liu:2022chg}, interfaced with {\sc Kira}. Their numerical evaluation, up to 50 digits, requires $\sim4.5$h on a laptop using 8 threads.
Finally, we have solved the system of differential equations in the Mandelstam variables $s$ and $t$ using the {\sc Mathematica} package {\sc SeaSyde}~\cite{Armadillo:2022ugh}, that allows us to consider complex-valued internal masses. 

Since when combining the numerical values of the MIs with their rational coefficients big numerical cancellations can occur, it is extremely important to keep under control the numerical precision of the evaluation of the MIs. 
Within {\sc SeaSyde}, we have decided to keep $70$ terms in the expansion in the kinematic invariants. The estimate for the relative error provided by {\sc SeaSyde} is, at worst, $10^{-14}$ in every point of the phase-space\footnote{The error is estimated by considering the relative contribution of the last $3$ terms in the series expansion over the complete series, both evaluated at half the radius of convergence.}. 
In several points of the phase-space we have performed a cross-check against {\sc AMFlow} by computing the MIs with $50$ digits precision, finding agreement for the first $14$ digits, in accordance to the uncertainty estimated by {\sc SeaSyde}. Cross-checks with analytical expressions, when available, have been performed as-well, finding agreement.

Using {\sc SeaSyde} we were able to generate a numerical grid in $(\sqrt{s},\cos\theta)$, consisting in $3250$ points, as described in detail in Section \ref{sec:numerics}.
The numerical evaluation of the complete grid for  all the topologies requires, in total, $\sim3$ weeks on a cluster with 26 cores. The most complicated case is the two-loop box with two massive internal lines, for which we have to solve a system with $56$ equations. The calculation of its grid requires $\sim10$ days by itself. 

\subsubsection{Mass evolution}
\label{sec:massevol}

\begin{figure}[t]
\begin{center}
\includegraphics[width=10cm]{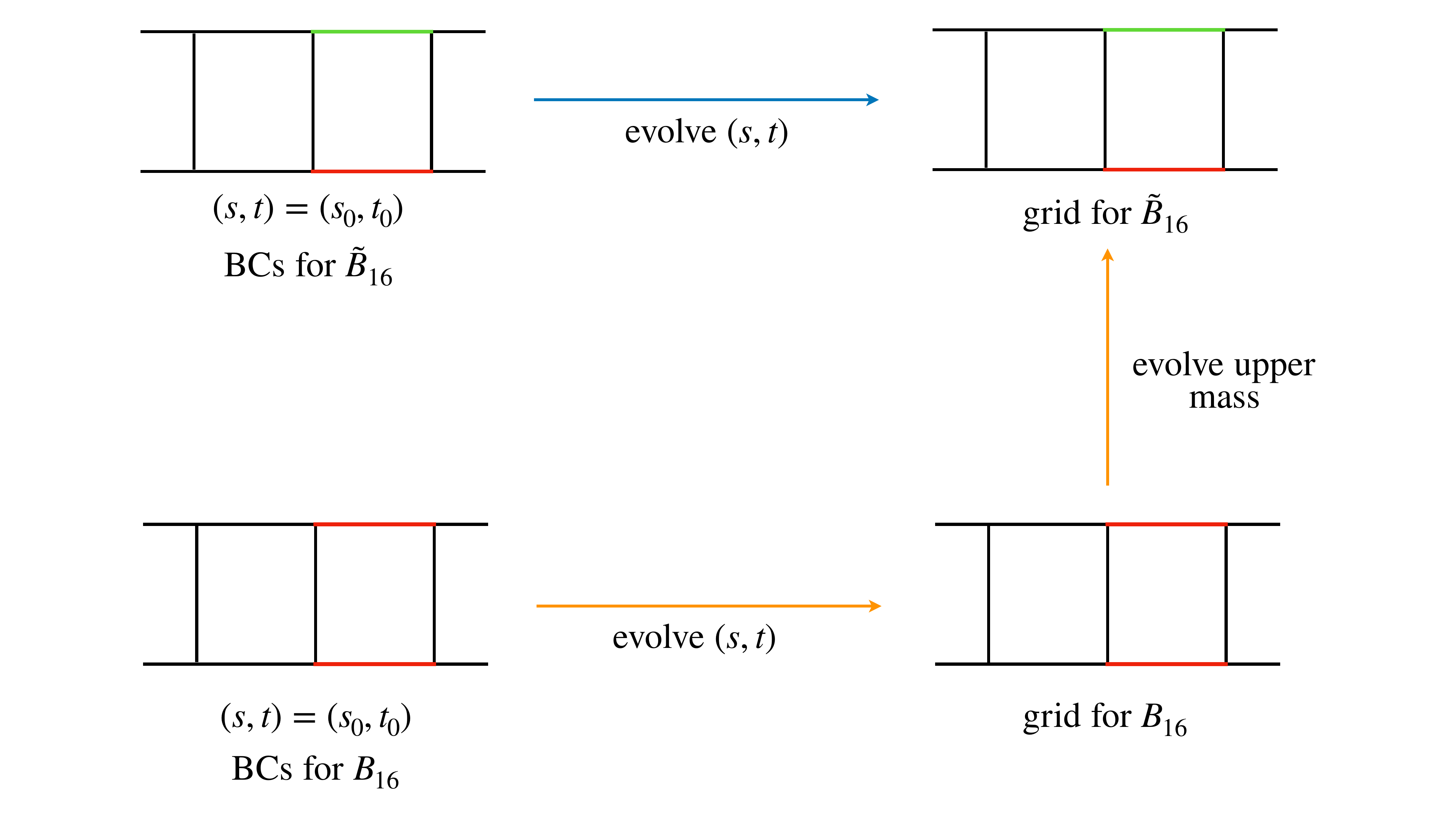}
\end{center}
\caption{
\label{fig:massevolution}
The two possibilities for creating a numerical grid for the $\tilde{B}_{16}$ topology. The first one, represented by the blue line, is to get the BCs for $\tilde{B}_{16}$ and then use the diff. eqs. w.r.t. $s$ and $t$ for obtaining the final grid. The second one, indicated by the orange lines, is to obtain the BCs for $B_{16}$, use the diff. eqs. for creating a grid in $s$ and $t$ for $B_{16}$ and, finally, use it as new BCs for evolving $B_{16}$ to $\tilde{B}_{16}$.
}
\end{figure}

Finally, we have done a self-consistency check by exploiting the flexibility given by the method of differential equations and the series expansion approach. We observe, indeed, that in the limit of the two masses being equal, the $\tilde{B}_{16}$ topology reduces to the $B_{16}$ from the neutral-current case. 
Hence, another way for computing a numerical grid for $\tilde{B}_{16}$ is, firstly, to create a numerical grid in $(s,t)$ for $B_{16}$ and, secondly, to write down the differential equations w.r.t. one of the two masses to evolve $B_{16}$ to $\tilde{B}_{16}$. We have explicitly verified, for different points in the phase-space, that the result obtained by evolving the equal masses box in $s$ and $ t$ and then in one of the two masses, is in agreement with the one obtained by evolving directly the two-masses box in the Mandelstam variables $s$ and $t$. The two possibilities are schematically depicted in Figure \ref{fig:massevolution}.

\subsection{Additional comments on the numerical evaluation of MIs}
\label{sec:additionalcomments}

In this work, our main objective is 
to pursue the automation of all the steps in the calculation of virtual corrections, in particular, the numerical evaluation of the MIs. Within our framework, all the steps are handled automatically (computing the differential equations, obtaining the boundary conditions and solving the system) thus opening the door to more complicated problems with the presence of a larger number of topologies.

However, this automation comes at a price. 
In \cite{Armadillo:2022bgm}, indeed, the computation of a numerical grid for the $B_{16}$ topology takes $\sim12$ hours. Even considering a factor $1.5/2$ related to the bigger size of the system ($56$ equations for $\tilde{B}_{16}$ against $36$ for $B_{16}$), we are a factor $10/15$ slower with respect to the case of NC DY. This big difference in time is mainly due to the state of the system of differential equations. In the previous case, in fact, we have put a strong effort in simplifying the system by choosing a suitable change of basis to cast the system in pre-canonical form. This simpler form of the system enormously speeds up the computation. On the contrary, in this work, we have taken the MIs suggested by {\sc Kira} and we have blindly written down their differential equations. This means that, other than standard algebraic simplifications, we have not performed any other optimization. Finding a change of basis which simplifies the system of differential equations is indeed important. However, it requires a big effort and a profound knowledge of the problem, while this blind approach relies only on computational power.

A second comment applies to the possibility of using {\sc AMFlow} for the evaluation of the MIs in all the points of the phase-space. The evaluation of all the MIs takes $\sim3\text{h} 15\text{m}$ per point, running on 8 cores and asking for a precision of $16$ digits. 
By using the same setup as the one used for 
{\sc SeaSyde}, the complete run would approximately take $130$ days\footnote{
The run-times have been estimated by using the latest public versions of each code.
}.

A final comment regards a severe limiting factor when working with numerical grids. Usually, all the input parameters are fixed to their numerical values and they cannot be modified without rerunning the final steps of the calculation. 
This requires a large amount of time. The series expansion approach, however, can be exploited for providing, with limited additional work, also the dependence on an additional variable e.g. $\delta M_W\equiv M_W-\bar{M}_{W}$, where $\bar{M}_{W}$ is a reference value for the mass of the $W$ boson. This can be done following the idea of the mass evolution introduced in Section \ref{sec:massevol}. In particular, for each topology, we use the grid in $(s,t)$, that we already have with a reference value $\bar{M}_W$, as a boundary condition, and we solve the differential equations w.r.t. $M_W$ around $\bar{M}_W$. By doing so, it is possible to provide a grid in $(s,t)$ where each point is a series in $\delta M_W$. Moreover, since the $\delta M_W$ we are interested in are of the order of $0.1$ GeV at most, only a small number of terms in the series is needed. Therefore, the solution can be computed in a small amount of time. For illustration, 
the evaluation of the $\delta M_W$ dependence for the $\tilde{B}_{16}$ topology, for all $3250$ points, has taken $\sim 1.5$ days, keeping $15$ terms in the series. The difference between the {\sc SeaSyde} solution, evaluated in $\delta M_W=0.1$ GeV, and {\sc AMFlow} is on the 13th significant digit. This approach opens the possibility for implementing this kind of corrections in frontier studies like, but not limited to, the $W$-boson mass determination.
\section{Infrared Singularities and Universal Pole Structure}

\label{sec:infra}
The bare result presented in Eq.~(\ref{eq:sumint}), thanks to the renormalisation procedure described in Section~\ref{sec:UV}, is free of divergences of UV origin.
Nevertheless, it still contains divergences of IR origin, generated by the exchange of soft and/or collinear massless partons.
While these singularities are guaranteed to cancel in the computation of an IR-safe observable, where the two-loop amplitude is combined with the corresponding real emission contributions, they are expected to appear at intermediate stages of a calculation.
Their presence requires the development of techniques to systematically handle and subtract them in a consistent way among the different contributions to an IR-safe observable.

At NLO, the Catani-Seymour dipole subtraction~\cite{Catani:1996jh,Catani:1996vz,Catani:2002hc} and FKS subtraction~\cite{Frixione:1995ms} are the two methods most widely used. 
In their complete generality, they can be applied to any process and they have been implemented in automatic routines in several computational frameworks.
At NNLO several techniques have been proposed (see e.g. \cite{Proceedings:2018jsb} and references therein) but none of them can yet claim full generality.
Regardless of the subtraction procedure, the IR poles are removed from the virtual contribution by using a process-independent subtraction operator. 
Such operators can in principle be different for each subtraction method but, because of the universal nature of the IR structure of the amplitude  \cite{Catani:1998bh,Sterman:2002qn,Becher:2009cu,Gardi:2009qi,Kilgore:2011pa,Kilgore:2013uta,
Becher:2007cu,Ahmed:2017gyt,Blumlein:2018tmz}, they can at most differ from each other by a finite contribution.

In this paper, we present as our final result the amplitudes after the subtraction of the IR divergences according to the $q_{\rm T}$ subtraction formalism \cite{Buonocore:2019puv,Buonocore:2021rxx}. 
We show them in the form of the hard function $H^{(1,1)}$, defined as the ratio of the 2-loop subtracted matrix element and the Born squared matrix element:
\begin{equation}
 H^{(1,1)} =
 \frac{1}{16}~
\left[
  2 ~ {\rm Re} \left( \frac{\langle \M^{(0)} | \M^{(1,1),fin} \rangle}{\langle \M^{(0)} | \M^{(0)} \rangle} \right)
\right]
  \;.
  \label{eq:Honeone}
\end{equation}

The computation of the IR-subtracted matrix element $\langle \M^{(0)} | \M^{(1,1),fin} \rangle$ requires the knowledge of the subtracted two-loop amplitude $| \M^{(1,1),fin} \rangle$.
We define it as follows:
\begin{equation}
  | \M^{(1,1),fin} \rangle =
  | \M^{(1,1)} \rangle -  \I^{(1,1)} | \M^{(0)} \rangle
                      - \tilde{\I}^{(0,1)} | \M^{(1,0),fin} \rangle
                      - \tilde{\I}^{(1,0)} | \M^{(0,1),fin} \rangle \, ,
 \label{eq:subtracted}
\end{equation}
where the $\mathcal I$s are the IR subtraction operators and $|\M^{(1,0),fin}\rangle$, $|\M^{(0,1),fin}\rangle$ are the finite reminders of the one-loop QCD and EW amplitudes respectively:
\begin{align}
 | \M^{(1,0),fin} \rangle &= | \M^{(1,0)} \rangle -  \I^{(1,0)} | \M^{(0)} \rangle \,,
 \label{eq:QCDfin}\\
 | \M^{(0,1),fin} \rangle &= | \M^{(0,1)} \rangle -  \I^{(0,1)} | \M^{(0)} \rangle \,.
 \label{eq:EWfin}
\end{align}

The subtraction operators can be obtained from the ones used in the case of the NC DY process after appropriately changing the charges of the initial state quarks and after neglecting the contribution stemming from the exchange of a photon between two final state particles, which is not present in the case of CC DY.
By indicating with $Q_i$ the value of the electric charge of the particle $i$ in 
positron units\footnote{E.g. $Q_u = \frac{2}{3}$.}, and with $C_F=\frac{N^2-1}{2 N}$ the Casimir of the fundamental representation of SU(N), the subtraction operators at one loop read:
\begin{align}
\I^{(1,0)} &= \smu^{-\ep} C_F \left( - \frac{2}{\ep^2} - \frac{1}{\ep} (3 + 2 i \pi)  + \zeta_2 \right) \,,
\\
\I^{(0,1)} &= \smu^{-\ep} \bigg[ \frac{Q_u^2+Q_d^2}{2} \left( - \frac{2}{\ep^2} - \frac{1}{\ep} (3 + 2 i \pi)  + \zeta_2 \right)
 + \frac{4}{\ep} {\Gamma}_{l}^{(0,1)} \bigg] \,,
\end{align}
where
\begin{align}
 {\Gamma}_{l}^{(0,1)} = 
 -\frac{1}{4}\bigg[
 Q_l^2 &\left(1-i\pi\right) 
 + Q_l^2 \log \left( \frac{m_l^2}{s} \right)+  \nonumber\\
 &+ 2 Q_u Q_l \log\left(\frac{(2 p_1\cdot p_4)}{s}\right)- 2 Q_d Q_l \log\left(\frac{(2 p_2\cdot p_4)}{s}\right)\bigg].
\end{align}

The two-loop subtraction operator for the mixed contribution reads:
\begin{align}
  \I^{(1,1)} = &\smu^{-2\ep} C_F 
\bigg[
\frac{Q_u^2+Q_d^2}{2}  \bigg( \frac{4}{\ep^4} + \frac{1}{\ep^3} ( 12 + 8 i \pi ) + \frac{1}{\ep^2} ( 9 - 28 \zeta_2 + 12 i \pi)+
\nonumber\\
&\;+ \frac{1}{\ep} \Big( -\frac{3}{2} + 6 \zeta_2- 24 \zeta_3 - 4 i \pi \zeta_2 \Big) \bigg)
+ \left( - \frac{2}{\ep^2} - \frac{1}{\ep} (3 + 2 i \pi)  + \zeta_2 \right)~\frac{4}{\ep}~\Gamma_l^{(0,1)} \bigg].
\label{eq:i11}
\end{align}
Following the same convention used in the case of the NC DY process, in Eq.~(\ref{eq:subtracted}) the subtraction of the one-loop-like divergences from the two loop amplitude is performed by using the  subtraction operators $\tilde{\I}^{(1,0)}$ and $\tilde{\I}^{(0,1)}$, which can be obtained from ${\I}^{(1,0)}$ and ${\I}^{(0,1)}$ by dropping
the term proportional to $\zeta_2$. 

The approximation of the amplitude in the small lepton mass limit
retains all the terms enhanced by $\log (m_l)$,
divergent in the $\ml\to 0$ limit.
The structure of these corrections reflects the universality property
of the final-state collinear divergences,
and is given, normalised to the Born squared matrix element, by
\begin{equation}
  \lim_{\ml\to0}
  \frac{\langle \M^{(0)} | \M^{(1,1),fin} \rangle}{\langle \M^{(0)} | \M^{(0)}\rangle}=
K~+~
 \frac{C_F}{2} Q_l^2 ( -8+7 \zeta_2 - 3 i \pi )
\bigg[ - \log \bigg( \frac{\ml^2}{s} \bigg)
       + \log^2  \bigg( \frac{\ml^2}{s} \bigg)
\bigg] \,,
\end{equation}

where $K$ represents all the other terms in the interference,
constant in the $\ml\to 0$ limit.
Note that the coefficients of the lepton mass logarithms are exactly 
with a factor half of those of the NC DY, indicating the universal
behaviour of these logarithms. 
\section{Results}
\label{sec:results}

\subsection{Numerical results}
\label{sec:numerics}
The evaluation of the finite IR-subtracted UV-renormalised hard function ${H}^{(1,1)}$,  defined in Eq. (\ref{eq:Honeone}),
requires the combination of several contributions, with a non-negligible evaluation time for the MIs.
For this reason it is of practical interest to prepare a numerical grid, which covers the whole phase space relevant in the applications at hadron colliders, making negligible the evaluation time of the results, at any arbitrary point.
We consider the partonic centre-of-mass energy $\sqrt{s}$ and scattering angle $\cos\theta$
and compute a grid with respectively 
(130x25) points,
covering the intervals $\sqrt{s}\in [40,8000]$ GeV and $\cos\theta\in [-1,1]$.
The sampling is based on the known behaviour
of the CC-DY NLO-EW distribution,
with special care for the $W$ resonance region,
where a finer binning is necessary.
We have verified 
that the interpolation describes the exact results with an accuracy, in the whole phase space, at least at the $10^{-3}$ level, guaranteed by the smoothness of the IR-subtracted ${H}^{(1,1)}$ function.
\begin{figure}[t]
\begin{center}
\begin{minipage}[t]{.475\textwidth}
\includegraphics[trim={0 0 0 2cm},clip, width=\textwidth]{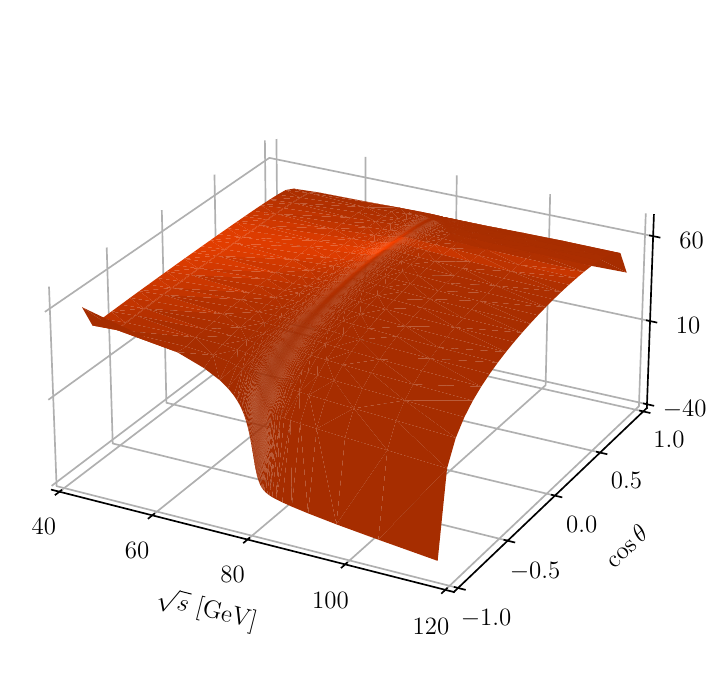}
\end{minipage}
\begin{minipage}[t]{.475\textwidth}
\includegraphics[trim={0 0.8cm 0 1cm},clip,width=\textwidth]{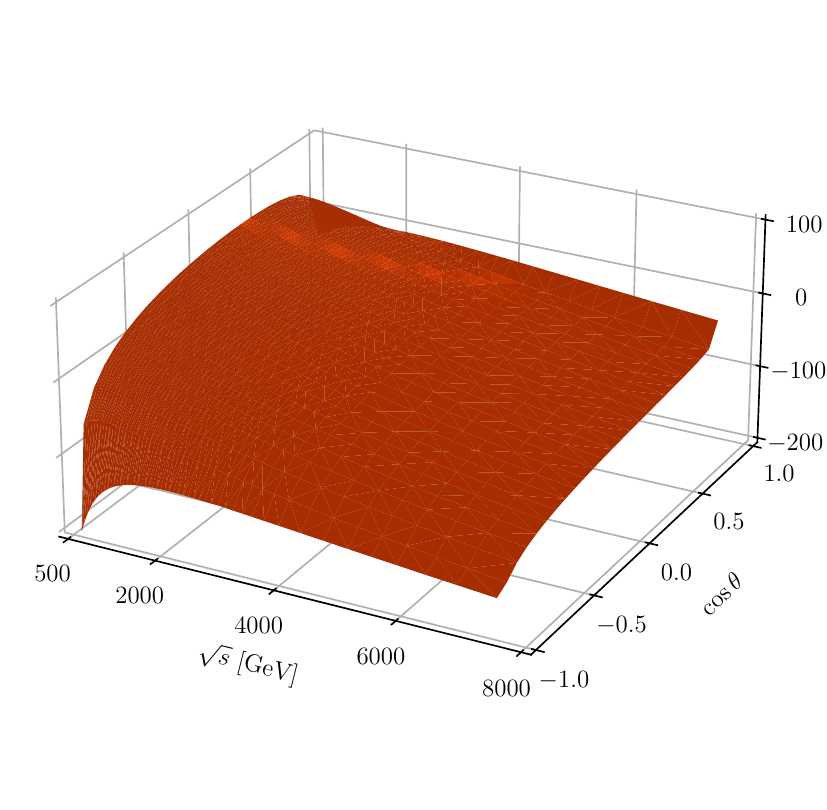}
\end{minipage}
\end{center}
\vspace{-2ex}
\caption{\label{fig:correction}
 The complete correction to the finite hard function in the $G_\mu$-scheme, due to \oaas correction,
  in two different phase-space regions,
  as a function of $\sqrt{s}$ and $\cos\theta$. The correction is normalized by the Born and expressed in units $\frac{\alpha}{\pi}\frac{\alpha_s}{\pi}$.
  }
\end{figure}

We present in Figure \ref{fig:correction}
the hard function $H^{(1,1)}$ in the $G_\mu$-scheme, which is normalised to the Born cross section
and expressed, as a function of $\sqrt{s}$ and $\cos\theta$, in units $\frac{\alpha}{\pi} \frac{\alpha_s}{\pi}$.
We consider 
for the partonic center-of-mass energy, two different intervals,
namely $40\leq\sqrt{s}\leq 120$ GeV and $500\leq\sqrt{s}\leq 7500$ GeV,
while the range in $\cos\theta$ is $[-1, 1]$.
In Figure \ref{fig:correction}, we use the following parameters:
\begin{center}
 \begin{tabular}{c l l l}
 \hline\hline
  $M_Z$ & 91.1535 GeV & $\Gamma_Z$ & 2.4943 GeV\\
  $M_W$ & 80.358 GeV  & $\Gamma_W$ & 2.084 GeV\\
  $m_H$ & 125.25 GeV  & $m_t$ & 173.2 GeV\\
 \hline\hline
 \end{tabular}
\end{center}
The possibility to have an exact dependence on the $W$-boson mass, within this numerical approach, is discussed in Section \ref{sec:Wexpansion}.

We provide in Table \ref{tab:benchmark} a few benchmark values of the function $H^{(1,1)}$ in the $G_\mu$-scheme, for different $\sqrt{s}$ and $\cos\theta$ choices.
\begin{table}
\begin{center}
 \begin{tabular}{c c c}
 \hline
$\sqrt{s}$ [GeV]  & $\cos\theta$ & $H^{(1,1)}$\\
 \hline\hline
  88.066    & 0     &  61.318  \\
  88.066    & -0.66 &  45.970  \\
  222.362   & 0     &  48.189  \\
  222.362   & -0.66 &  -15.753 \\
  1035.37   & 0     &  29.029  \\
  1035.37   & -0.66 &  -6.990  \\
 \hline
 \end{tabular}
\caption{\label{tab:benchmark}
 Benchmark values for the finite hard function, in the $G_\mu$-scheme, due to \oaas correction.
 The correction is normalized by the Born and expressed in units $\frac{\alpha}{\pi}\frac{\alpha_s}{\pi}$.
  }
\end{center}
\end{table}

\subsection{Checks}
The scattering amplitude develops UV and IR divergences, which appear as poles  in the dimensional regularisation parameter $\varepsilon$. Their cancellation provides a non trivial check of the consistency of the calculation. This check exploits the restoration of some QED-like Ward identities, valid in the BFG.
In the construction of the UV-finite renormalised propagator we observe the cancellation of the $W$ self-energy wave function divergence against the one of the charge counterterms.
We verify that the sum of external wave function factors, vertex and box corrections is UV finite, thanks to the validity of the above mentioned Ward identities, featuring only IR singularities.
The latter have indeed a universal structure,  leading to the independent construction of an IR subtraction term, presented in Section \ref{sec:infra}.
The singularities of this subtraction term exactly match those of the sum of wave function, vertex and box corrections, leaving a finite remainder, free of any divergence.

The cancellation is not trivial, because in the evaluation of the amplitude, IR subtraction term, and UV counterterms we need to specify a prescription to handle the Dirac's $\gamma_5$ matrix in $d$ dimensions, yielding additional poles in $\varepsilon$.
Each contribution is in fact prescription dependent, but their sum is not, as confirmed by the complete cancellation of the divergences.
In several stages of our computations and checks, we have used
{\sc GiNaC} \cite{Vollinga:2004sn}, {\sc HarmonicSums} \cite{Ablinger:2010kw,Ablinger:2014rba}, 
{\sc PolyLogTools} \cite{Duhr:2019tlz}  and {\sc LoopTools} \cite{Hahn:1998yk}.

\subsection{Full dependence on the \texorpdfstring{$W$}{W}-boson mass in the numerical grids}
\label{sec:Wexpansion}
In Sections \ref{sec:semianalytical} and \ref{sec:additionalcomments} we have introduced the idea of using the series expansion approach in order to provide a grid in $(s,t)$ in which each point is a series expansion in $\delta M_W$, thus featuring the exact $\mu_W$ dependence.
In this Section we discuss the sensitivity to a parameter like the $W$-boson mass and the accuracy of a series expansion approach compared to the exact evaluation. 

\begin{figure}[t]
\begin{center}
\begin{minipage}[t]{.475\textwidth}
\includegraphics[width=\textwidth]{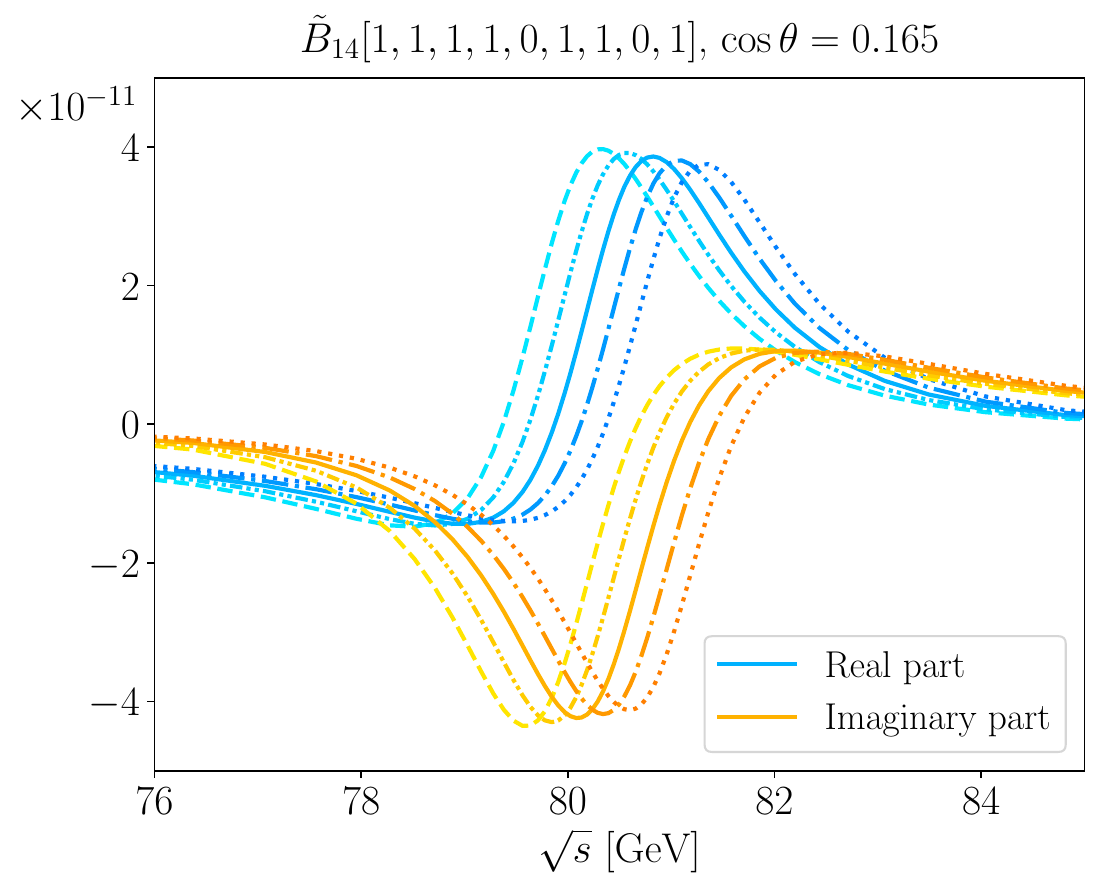}
\end{minipage}
\begin{minipage}[t]{.495\textwidth}
\includegraphics[width=\textwidth]{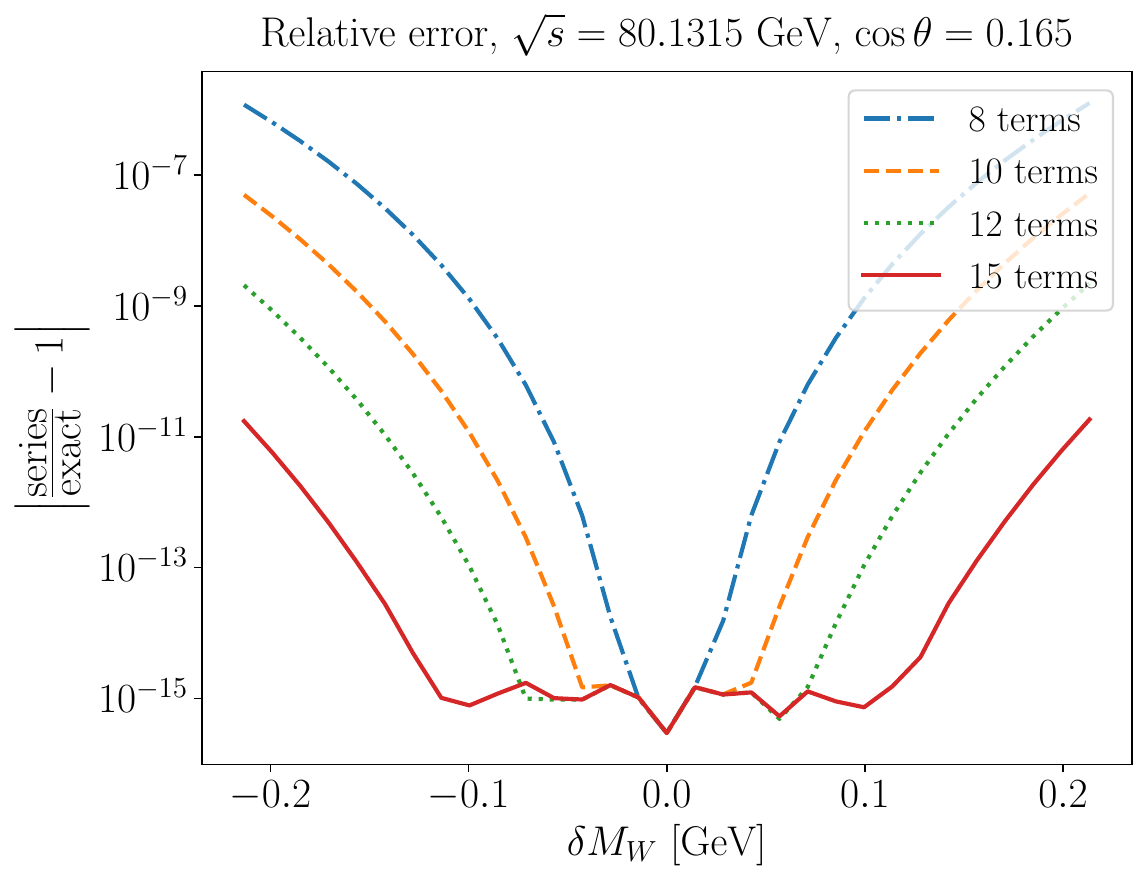}
\end{minipage}
\end{center}
\caption{\label{fig:dmw}
On the left panel we plot the real and imaginary part of the $\mathcal{O}(\varepsilon^0)$ of $\tilde{B}_{14}[1,1,1,1,0,1,1,0,1]$ for different values of $\delta M_W$. 
On the right panel we plot the relative error of the solution for different number of terms in the $\delta M_W$-series expansion, as a function of $\delta M_W$.
}
\end{figure}

As an example, we consider the integral $\tilde{B}_{14}[1,1,1,1,0,1,1,0,1]$, which is defined as:
\be
\int\frac{d^d k_1}{(2\pi)^d}\frac{d^d k_2}{(2\pi)^d}\frac{1}{
\cD_1\;                 
(\cD_2-\mu_W^2)\;        
\cD_{12}\;              
\cD_{1;1}\;             
\cD_{2;1}^0\;           
\cD_{1;12}\;            
\cD_{2;12}\;            
\cD_{1;3}^0\;           
(\cD_{2;3}-m_\ell^2)}   
\ee
with the $\cD$s introduced in Eq.~(\ref{eq:denominators}). This integral can be represented graphically as
\begin{figure}[H]
\centering
\includegraphics[trim={2cm 5cm 0 5cm},clip,width=0.35\textwidth]{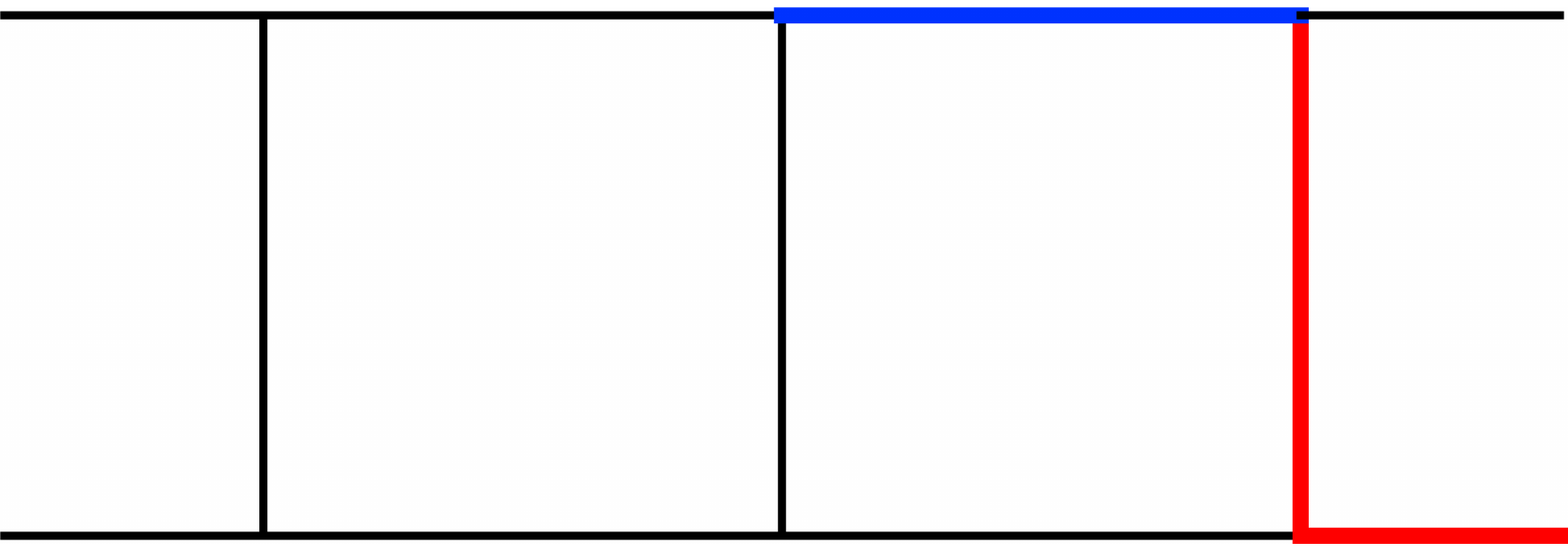}
\end{figure}
\noindent
where the blue (red) indicates that a particle with mass $\mu_W$ ($m_\ell$) is running in the line, and it appears, for example, in diagrams with the exchange of a photon and a $W$ boson.
In the left panel of Figure \ref{fig:dmw} we plot the real and imaginary part of the integral, for different values of $\mw$, as a function of $\sqrt{s}$, with $\cos\theta=0.165$. The intensity of the color corresponds to 
different choices of $\delta M_W$, from $-500$ MeV to $500$ MeV, in steps of $250$ MeV.
The shift of the peak position illustrates the sensitivity of this integral to the choice of the mass value.
In the right panel of Figure \ref{fig:dmw} we compare the exact evaluation of the integral\footnote{The exact solution is obtained using {\sc AMFlow}, asking for a precision of 30 digits.} against different approximations. We consider $\sqrt{s}=80.1315$ GeV in order to have a strong dependence on $\mw$ and $\cos\theta=0.165$. The curves correspond to a different number of terms in the $\delta M_W$ expansion, and their ratio with the exact result is plotted as a function of $\delta M_W$. From the plot we can see that if we consider only shifts in $\delta M_W$ of order 100 MeV, 15 terms in the expansions are sufficient for maintaining a relative precision of $10^{-15}$.

\section{Conclusions}

\label{sec:conclusions}
We have presented in this paper the details of the complete calculation, for the CC DY process, 
of its exact \oaas two-loop virtual corrections.
These results represent the companion to the ones discussed in Ref. \cite{Armadillo:2022bgm} for the NC DY case, with a higher level of technical complexity in the Master Integrals, because of the presence of two different mass values in the internal lines.
When included in the {\sc Matrix} framework, for the evaluation of the fiducial cross sections, these results will allow a consistent simultaneous analysis of both NC and CC DY processes at NNLO QCD-EW level. Such consistency is required by the interplay between the two final states: for example, in the $W$-boson mass studies the NC DY channel plays a crucial calibration role, which would be spoiled if corrections at different orders were considered;  at large lepton-pair transverse/invariant masses, CC and NC channels have different sensitivity to the parton-parton luminosities, thus allowing an effective reduction of the associated uncertainties, crucial in the New Physics searches.

The results have been obtained thanks to an increased level of automation of every step of the calculation, opening the way to the systematic study of the mixed QCD-EW corrections in other $2\to 2$ scattering processes.
In particular, it is worth mentioning the possibility to study in a uniform way all the relevant MIs, with 0,1, or 2 internal massive lines, in the same semi-analytical framework offered by the {\sc SeaSyde} code, with excellent control on the cancellation of UV and IR divergences.

The flexibility of the differential equations technique to solve the MIs has been exploited to preserve the exact dependence on the $W$-boson mass, even when we prepare the numerical grid for the total correction factor. We achieve in this way excellent performances in the numerical evaluation together with full control on the accuracy of the result.

\section*{Acknowledgments}
We would like to thank L. Buonocore, M. Grazzini, S. Kallweit, C. Savoini and F. Tramontano,
for several interesting discussions during the development of the complete calculation of the NNLO QCD-EW corrections to NC DY and for a careful reading of the manuscript.
T.A. is a Research Fellow of the Fonds de la Recherche Scientifique – FNRS.
The work of S.D. has been partially funded by the European Union (ERC, MultiScaleAmp, Grant Agreement No. 101078449). Views and opinions expressed are however those of the author(s) only and do not necessarily reflect those of the European Union or the European Research Council Executive Agency. Neither the European Union nor the granting authority can be held responsible for them.

\appendix
\section{Description of the ancillary files}
\label{app:readme}
In this appendix we describe the ancillary {\sc Mathematica} file \textsf{CCDY\_M11.m}.
This file contains the expression for $\langle \hat\M^{(0)} | \hat\M^{(1,1)} \rangle$ i.e.
the unrenormalised interference term between the Born and the bare 2-loop amplitude as defined in Eq.~\ref{eq:sumint}. In particular, within $|\hat\M^{(1,1)}\rangle$, there are all the pure 2-loop contribution, which are schematically depicted in Figure \ref{fig:diagsamples}-(a), (c), (f) and \ref{fig:diagsamplesCT}-(a), (c).

The result is provided as a sum of rational coefficients times a MI:
\be
\sum_k c_k (s,t,\{m_i\},d) \; I_k (s,t,\{m_i\},d).
\ee
In the file, we make use of the following symbols:
\begin{align}
&\mathsf{mz}=\mz \;, \qquad
\mathsf{mzC}={\mu}_Z^{*} \;, \qquad
\mathsf{mw}=\mw \;, \qquad
\mathsf{mwC}={\mu}_W^{*} \;, \nonumber \\
&\mathsf{sw}=s_W \;, \qquad
\mathsf{swC}=s_W^{*} \;, \qquad
\mathsf{cw}=c_W \;, \qquad
\mathsf{cwC}={c}_W^{*} \;, \nonumber \\
&\mathsf{mm}=m_\ell \;, \qquad 
\mathsf{mt}=m_t \;, \qquad 
\mathsf{mh}=m_H \;, \qquad 
\mathsf{d}=4-2\varepsilon \nonumber \\
&\mathsf{prefactor}=\left(\frac{\alpha}{4\pi}\right)^2 (4\pi)^8 \mu^{8-2d}.
\end{align}
where $s_W$ and $c_W$ are the sine and cosine of the Weinberg angle, the $^*$ indicates the complex conjugate and $\mu$ is the renormalisation scale.

In the file, we label each MI as :
\be
\mathsf{Topology[\{masses\_\_\}, x\_\_]}.
\ee
where $\mathsf{Topology}$ is the name of the topology, $\mathsf{\{masses\_\_\}}$ is a list of the masses in the same order as they appear in the topology definition and $\mathsf{x\_\_}$ is the list of powers to which each denominator is raised. Additionally, all the integrals have a normalization factor $1/(2\pi)^d$ for each loop. 

For example, $\mathsf{B16tilde[\{mw, mz\}, 0, 1, 1, 0, 0, 1, 1, 0, 0]} $ is a short notation for:
\be
\int\frac{d^d k_1}{(2\pi)^d}\frac{d^d k_2}{(2\pi)^d}
\frac{1}{(\cD_2-\mw^2)\cD_{12}\cD_{1;12}(\cD_{2;12}-\mz^2)}.
\ee

We provide also the file $\mathsf{integralfamilies.yaml}$ with all the integral families expressed in a format suitable for a reduction with {\sc Kira}.

\bibliography{long}

\providecommand{\href}[2]{#2}\begingroup\raggedright\begin{thebibliography}{100}

\bibitem{Drell:1970wh}
S.~Drell and T.-M. Yan, \emph{{Massive Lepton Pair Production in Hadron-Hadron Collisions at High-Energies}}, \href{https://doi.org/10.1103/PhysRevLett.25.316}{\emph{Phys. Rev. Lett.} {\bfseries 25} (1970) 316--320}.

\bibitem{Group:2012gb}
{\scshape CDF, D0} collaboration, T.~E.~W. Group, \emph{{2012 Update of the Combination of CDF and D0 Results for the Mass of the W Boson}},  \href{https://arxiv.org/abs/1204.0042}{{\ttfamily 1204.0042}}.

\bibitem{Aaboud:2017svj}
{\scshape ATLAS} collaboration, M.~Aaboud et~al., \emph{{Measurement of the $W$-boson mass in pp collisions at $\sqrt{s}=7$ TeV with the ATLAS detector}}, \href{https://doi.org/10.1140/epjc/s10052-017-5475-4}{\emph{Eur. Phys. J. C} {\bfseries 78} (2018) 110}, [\href{https://arxiv.org/abs/1701.07240}{{\ttfamily 1701.07240}}].

\bibitem{Aaltonen:2018dxj}
{\scshape CDF, D0} collaboration, T.~A. Aaltonen et~al., \emph{{Tevatron Run II combination of the effective leptonic electroweak mixing angle}}, \href{https://doi.org/10.1103/PhysRevD.97.112007}{\emph{Phys. Rev. D} {\bfseries 97} (2018) 112007}, [\href{https://arxiv.org/abs/1801.06283}{{\ttfamily 1801.06283}}].

\bibitem{ATLAS:2018gqq}
{\scshape ATLAS} collaboration, \emph{{Measurement of the effective leptonic weak mixing angle using electron and muon pairs from $Z$-boson decay in the ATLAS experiment at $\sqrt s = 8$ TeV}}, {\emph{ATLAS-CONF-2018-037} (7, 2018) }.

\bibitem{CMS-PAS-SMP-22-010}
{\scshape CMS} collaboration, \emph{{Measurement of the Drell-Yan forward-backward asymmetry and of the effective leptonic weak mixing angle using proton-proton collisions at 13 TeV}},  tech. rep., CERN, Geneva, 2024.

\bibitem{CarloniCalame:2016ouw}
C.~M. Carloni~Calame, M.~Chiesa, H.~Martinez, G.~Montagna, O.~Nicrosini, F.~Piccinini and A.~Vicini, \emph{{Precision Measurement of the W-Boson Mass: Theoretical Contributions and Uncertainties}}, \href{https://doi.org/10.1103/PhysRevD.96.093005}{\emph{Phys. Rev. D} {\bfseries 96} (2017) 093005}, [\href{https://arxiv.org/abs/1612.02841}{{\ttfamily 1612.02841}}].

\bibitem{Bagnaschi:2019mzi}
E.~Bagnaschi and A.~Vicini, \emph{{Parton Density Uncertainties and the Determination of Electroweak Parameters at Hadron Colliders}}, \href{https://doi.org/10.1103/PhysRevLett.126.041801}{\emph{Phys. Rev. Lett.} {\bfseries 126} (2021) 041801}, [\href{https://arxiv.org/abs/1910.04726}{{\ttfamily 1910.04726}}].

\bibitem{Behring:2021adr}
A.~Behring, F.~Buccioni, F.~Caola, M.~Delto, M.~Jaquier, K.~Melnikov and R.~R\"ontsch, \emph{{Estimating the impact of mixed QCD-electroweak corrections on the $W$-mass determination at the LHC}}, \href{https://doi.org/10.1103/PhysRevD.103.113002}{\emph{Phys. Rev. D} {\bfseries 103} (2021) 113002}, [\href{https://arxiv.org/abs/2103.02671}{{\ttfamily 2103.02671}}].

\bibitem{Rottoli:2023xdc}
L.~Rottoli, P.~Torrielli and A.~Vicini, \emph{{Determination of the W-boson mass at hadron colliders}}, \href{https://doi.org/10.1140/epjc/s10052-023-12128-z}{\emph{Eur. Phys. J. C} {\bfseries 83} (2023) 948}, [\href{https://arxiv.org/abs/2301.04059}{{\ttfamily 2301.04059}}].

\bibitem{Torrielli:2023tiz}
P.~Torrielli, L.~Rottoli and A.~Vicini, \emph{{A new observable for W-mass determination}}, \href{https://doi.org/10.22323/1.432.0038}{\emph{PoS} {\bfseries RADCOR2023} (2024) 038}, [\href{https://arxiv.org/abs/2308.15993}{{\ttfamily 2308.15993}}].

\bibitem{Altarelli:1979ub}
G.~Altarelli, R.~Ellis and G.~Martinelli, \emph{{Large Perturbative Corrections to the Drell-Yan Process in QCD}}, \href{https://doi.org/10.1016/0550-3213(79)90116-0}{\emph{Nucl. Phys. B} {\bfseries 157} (1979) 461--497}.

\bibitem{Hamberg:1990np}
R.~Hamberg, W.~van Neerven and T.~Matsuura, \emph{{A complete calculation of the order $\alpha_s^{2}$ correction to the Drell-Yan $K$ factor}}, \href{https://doi.org/10.1016/0550-3213(91)90064-5}{\emph{Nucl. Phys. B} {\bfseries 359} (1991) 343--405}.

\bibitem{Harlander:2002wh}
R.~V. Harlander and W.~B. Kilgore, \emph{{Next-to-next-to-leading order Higgs production at hadron colliders}}, \href{https://doi.org/10.1103/PhysRevLett.88.201801}{\emph{Phys. Rev. Lett.} {\bfseries 88} (2002) 201801}, [\href{https://arxiv.org/abs/hep-ph/0201206}{{\ttfamily hep-ph/0201206}}].

\bibitem{Duhr:2020seh}
C.~Duhr, F.~Dulat and B.~Mistlberger, \emph{{Drell-Yan Cross Section to Third Order in the Strong Coupling Constant}}, \href{https://doi.org/10.1103/PhysRevLett.125.172001}{\emph{Phys. Rev. Lett.} {\bfseries 125} (2020) 172001}, [\href{https://arxiv.org/abs/2001.07717}{{\ttfamily 2001.07717}}].

\bibitem{Duhr:2020sdp}
C.~Duhr, F.~Dulat and B.~Mistlberger, \emph{{Charged current Drell-Yan production at N$^{3}$LO}}, \href{https://doi.org/10.1007/JHEP11(2020)143}{\emph{JHEP} {\bfseries 11} (2020) 143}, [\href{https://arxiv.org/abs/2007.13313}{{\ttfamily 2007.13313}}].

\bibitem{Duhr:2021vwj}
C.~Duhr and B.~Mistlberger, \emph{{Lepton-pair production at hadron colliders at N$^{3}$LO in QCD}}, \href{https://doi.org/10.1007/JHEP03(2022)116}{\emph{JHEP} {\bfseries 03} (2022) 116}, [\href{https://arxiv.org/abs/2111.10379}{{\ttfamily 2111.10379}}].

\bibitem{Anastasiou:2003yy}
C.~Anastasiou, L.~J. Dixon, K.~Melnikov and F.~Petriello, \emph{{Dilepton rapidity distribution in the Drell-Yan process at NNLO in QCD}}, \href{https://doi.org/10.1103/PhysRevLett.91.182002}{\emph{Phys. Rev. Lett.} {\bfseries 91} (2003) 182002}, [\href{https://arxiv.org/abs/hep-ph/0306192}{{\ttfamily hep-ph/0306192}}].

\bibitem{Anastasiou:2003ds}
C.~Anastasiou, L.~J. Dixon, K.~Melnikov and F.~Petriello, \emph{{High precision QCD at hadron colliders: Electroweak gauge boson rapidity distributions at NNLO}}, \href{https://doi.org/10.1103/PhysRevD.69.094008}{\emph{Phys. Rev. D} {\bfseries 69} (2004) 094008}, [\href{https://arxiv.org/abs/hep-ph/0312266}{{\ttfamily hep-ph/0312266}}].

\bibitem{Melnikov:2006kv}
K.~Melnikov and F.~Petriello, \emph{{Electroweak gauge boson production at hadron colliders through $\mathcal{O}(\alpha_s^2)$}}, \href{https://doi.org/10.1103/PhysRevD.74.114017}{\emph{Phys. Rev. D} {\bfseries 74} (2006) 114017}, [\href{https://arxiv.org/abs/hep-ph/0609070}{{\ttfamily hep-ph/0609070}}].

\bibitem{Catani:2009sm}
S.~Catani, L.~Cieri, G.~Ferrera, D.~de~Florian and M.~Grazzini, \emph{{Vector boson production at hadron colliders: a fully exclusive QCD calculation at NNLO}}, \href{https://doi.org/10.1103/PhysRevLett.103.082001}{\emph{Phys. Rev. Lett.} {\bfseries 103} (2009) 082001}, [\href{https://arxiv.org/abs/0903.2120}{{\ttfamily 0903.2120}}].

\bibitem{Catani:2010en}
S.~Catani, G.~Ferrera and M.~Grazzini, \emph{{W Boson Production at Hadron Colliders: The Lepton Charge Asymmetry in NNLO QCD}}, \href{https://doi.org/10.1007/JHEP05(2010)006}{\emph{JHEP} {\bfseries 05} (2010) 006}, [\href{https://arxiv.org/abs/1002.3115}{{\ttfamily 1002.3115}}].

\bibitem{Camarda:2021ict}
S.~Camarda, L.~Cieri and G.~Ferrera, \emph{{Drell\textendash{}Yan lepton-pair production: qT resummation at N3LL accuracy and fiducial cross sections at N3LO}}, \href{https://doi.org/10.1103/PhysRevD.104.L111503}{\emph{Phys. Rev. D} {\bfseries 104} (2021) L111503}, [\href{https://arxiv.org/abs/2103.04974}{{\ttfamily 2103.04974}}].

\bibitem{Chen:2022cgv}
X.~Chen, T.~Gehrmann, E.~W.~N. Glover, A.~Huss, P.~F. Monni, E.~Re, L.~Rottoli and P.~Torrielli, \emph{{Third-Order Fiducial Predictions for Drell-Yan Production at the LHC}}, \href{https://doi.org/10.1103/PhysRevLett.128.252001}{\emph{Phys. Rev. Lett.} {\bfseries 128} (2022) 252001}, [\href{https://arxiv.org/abs/2203.01565}{{\ttfamily 2203.01565}}].

\bibitem{Neumann:2022lft}
T.~Neumann and J.~Campbell, \emph{{Fiducial Drell-Yan production at the LHC improved by transverse-momentum resummation at N4LLp+N3LO}}, \href{https://doi.org/10.1103/PhysRevD.107.L011506}{\emph{Phys. Rev. D} {\bfseries 107} (2023) L011506}, [\href{https://arxiv.org/abs/2207.07056}{{\ttfamily 2207.07056}}].

\bibitem{Campbell:2023lcy}
J.~Campbell and T.~Neumann, \emph{{Third order QCD predictions for fiducial W-boson production}}, \href{https://doi.org/10.1007/JHEP11(2023)127}{\emph{JHEP} {\bfseries 11} (2023) 127}, [\href{https://arxiv.org/abs/2308.15382}{{\ttfamily 2308.15382}}].

\bibitem{Chen:2021vtu}
X.~Chen, T.~Gehrmann, N.~Glover, A.~Huss, T.-Z. Yang and H.~X. Zhu, \emph{{Dilepton Rapidity Distribution in Drell-Yan Production to Third Order in QCD}}, \href{https://doi.org/10.1103/PhysRevLett.128.052001}{\emph{Phys. Rev. Lett.} {\bfseries 128} (2022) 052001}, [\href{https://arxiv.org/abs/2107.09085}{{\ttfamily 2107.09085}}].

\bibitem{Chen:2022lwc}
X.~Chen, T.~Gehrmann, N.~Glover, A.~Huss, T.-Z. Yang and H.~X. Zhu, \emph{{Transverse mass distribution and charge asymmetry in W boson production to third order in QCD}}, \href{https://doi.org/10.1016/j.physletb.2023.137876}{\emph{Phys. Lett. B} {\bfseries 840} (2023) 137876}, [\href{https://arxiv.org/abs/2205.11426}{{\ttfamily 2205.11426}}].

\bibitem{Moch:2005ky}
S.~Moch and A.~Vogt, \emph{{Higher-order soft corrections to lepton pair and Higgs boson production}}, \href{https://doi.org/10.1016/j.physletb.2005.09.061}{\emph{Phys. Lett. B} {\bfseries 631} (2005) 48--57}, [\href{https://arxiv.org/abs/hep-ph/0508265}{{\ttfamily hep-ph/0508265}}].

\bibitem{Laenen:2005uz}
E.~Laenen and L.~Magnea, \emph{{Threshold resummation for electroweak annihilation from DIS data}}, \href{https://doi.org/10.1016/j.physletb.2005.10.038}{\emph{Phys. Lett. B} {\bfseries 632} (2006) 270--276}, [\href{https://arxiv.org/abs/hep-ph/0508284}{{\ttfamily hep-ph/0508284}}].

\bibitem{Ravindran:2005vv}
V.~Ravindran, \emph{{On Sudakov and soft resummations in QCD}}, \href{https://doi.org/10.1016/j.nuclphysb.2006.04.008}{\emph{Nucl. Phys. B} {\bfseries 746} (2006) 58--76}, [\href{https://arxiv.org/abs/hep-ph/0512249}{{\ttfamily hep-ph/0512249}}].

\bibitem{Ravindran:2006cg}
V.~Ravindran, \emph{{Higher-order threshold effects to inclusive processes in QCD}}, \href{https://doi.org/10.1016/j.nuclphysb.2006.06.025}{\emph{Nucl. Phys. B} {\bfseries 752} (2006) 173--196}, [\href{https://arxiv.org/abs/hep-ph/0603041}{{\ttfamily hep-ph/0603041}}].

\bibitem{deFlorian:2012za}
D.~de~Florian and J.~Mazzitelli, \emph{{A next-to-next-to-leading order calculation of soft-virtual cross sections}}, \href{https://doi.org/10.1007/JHEP12(2012)08}{\emph{JHEP} {\bfseries 12} (2012) 088}, [\href{https://arxiv.org/abs/1209.0673}{{\ttfamily 1209.0673}}].

\bibitem{Ahmed:2014cla}
T.~Ahmed, M.~Mahakhud, N.~Rana and V.~Ravindran, \emph{{Drell-Yan Production at Threshold to Third Order in QCD}}, \href{https://doi.org/10.1103/PhysRevLett.113.112002}{\emph{Phys. Rev. Lett.} {\bfseries 113} (2014) 112002}, [\href{https://arxiv.org/abs/1404.0366}{{\ttfamily 1404.0366}}].

\bibitem{Catani:2014uta}
S.~Catani, L.~Cieri, D.~de~Florian, G.~Ferrera and M.~Grazzini, \emph{{Threshold resummation at N$^3$LL accuracy and soft-virtual cross sections at N$^3$LO}}, \href{https://doi.org/10.1016/j.nuclphysb.2014.09.012}{\emph{Nucl. Phys. B} {\bfseries 888} (2014) 75--91}, [\href{https://arxiv.org/abs/1405.4827}{{\ttfamily 1405.4827}}].

\bibitem{Li:2014afw}
Y.~Li, A.~von Manteuffel, R.~M. Schabinger and H.~X. Zhu, \emph{{Soft-virtual corrections to Higgs production at N$^3$LO}}, \href{https://doi.org/10.1103/PhysRevD.91.036008}{\emph{Phys. Rev. D} {\bfseries 91} (2015) 036008}, [\href{https://arxiv.org/abs/1412.2771}{{\ttfamily 1412.2771}}].

\bibitem{Ajjath:2020ulr}
A.~H. Ajjath, P.~Mukherjee and V.~Ravindran, \emph{{On next to soft corrections to Drell-Yan and Higgs Boson productions}},  \href{https://arxiv.org/abs/2006.06726}{{\ttfamily 2006.06726}}.

\bibitem{Dittmaier:2001ay}
S.~Dittmaier and M.~Kr\"amer, \emph{{Electroweak radiative corrections to W boson production at hadron colliders}}, \href{https://doi.org/10.1103/PhysRevD.65.073007}{\emph{Phys. Rev. D} {\bfseries 65} (2002) 073007}, [\href{https://arxiv.org/abs/hep-ph/0109062}{{\ttfamily hep-ph/0109062}}].

\bibitem{Baur:2004ig}
U.~Baur and D.~Wackeroth, \emph{{Electroweak radiative corrections to $p \bar{p} \to W^\pm \to \ell^\pm \nu$ beyond the pole approximation}}, \href{https://doi.org/10.1103/PhysRevD.70.073015}{\emph{Phys. Rev. D} {\bfseries 70} (2004) 073015}, [\href{https://arxiv.org/abs/hep-ph/0405191}{{\ttfamily hep-ph/0405191}}].

\bibitem{Zykunov:2006yb}
V.~Zykunov, \emph{{Radiative corrections to the Drell-Yan process at large dilepton invariant masses}}, \href{https://doi.org/10.1134/S1063778806090109}{\emph{Phys. Atom. Nucl.} {\bfseries 69} (2006) 1522}.

\bibitem{Arbuzov:2005dd}
A.~Arbuzov, D.~Bardin, S.~Bondarenko, P.~Christova, L.~Kalinovskaya, G.~Nanava and R.~Sadykov, \emph{{One-loop corrections to the Drell-Yan process in SANC. I. The Charged current case}}, \href{https://doi.org/10.1140/epjc/s2006-02505-y}{\emph{Eur. Phys. J. C} {\bfseries 46} (2006) 407--412}, [\href{https://arxiv.org/abs/hep-ph/0506110}{{\ttfamily hep-ph/0506110}}].

\bibitem{CarloniCalame:2006zq}
C.~Carloni~Calame, G.~Montagna, O.~Nicrosini and A.~Vicini, \emph{{Precision electroweak calculation of the charged current Drell-Yan process}}, \href{https://doi.org/10.1088/1126-6708/2006/12/016}{\emph{JHEP} {\bfseries 12} (2006) 016}, [\href{https://arxiv.org/abs/hep-ph/0609170}{{\ttfamily hep-ph/0609170}}].

\bibitem{Baur:2001ze}
U.~Baur, O.~Brein, W.~Hollik, C.~Schappacher and D.~Wackeroth, \emph{{Electroweak radiative corrections to neutral current Drell-Yan processes at hadron colliders}}, \href{https://doi.org/10.1103/PhysRevD.65.033007}{\emph{Phys. Rev. D} {\bfseries 65} (2002) 033007}, [\href{https://arxiv.org/abs/hep-ph/0108274}{{\ttfamily hep-ph/0108274}}].

\bibitem{Zykunov:2005tc}
V.~Zykunov, \emph{{Weak radiative corrections to Drell-Yan process for large invariant mass of di-lepton pair}}, \href{https://doi.org/10.1103/PhysRevD.75.073019}{\emph{Phys. Rev. D} {\bfseries 75} (2007) 073019}, [\href{https://arxiv.org/abs/hep-ph/0509315}{{\ttfamily hep-ph/0509315}}].

\bibitem{CarloniCalame:2007cd}
C.~Carloni~Calame, G.~Montagna, O.~Nicrosini and A.~Vicini, \emph{{Precision electroweak calculation of the production of a high transverse-momentum lepton pair at hadron colliders}}, \href{https://doi.org/10.1088/1126-6708/2007/10/109}{\emph{JHEP} {\bfseries 10} (2007) 109}, [\href{https://arxiv.org/abs/0710.1722}{{\ttfamily 0710.1722}}].

\bibitem{Arbuzov:2007db}
A.~Arbuzov, D.~Bardin, S.~Bondarenko, P.~Christova, L.~Kalinovskaya, G.~Nanava and R.~Sadykov, \emph{{One-loop corrections to the Drell--Yan process in SANC. (II). The Neutral current case}}, \href{https://doi.org/10.1140/epjc/s10052-008-0531-8}{\emph{Eur. Phys. J. C} {\bfseries 54} (2008) 451--460}, [\href{https://arxiv.org/abs/0711.0625}{{\ttfamily 0711.0625}}].

\bibitem{Dittmaier:2009cr}
S.~Dittmaier and M.~Huber, \emph{{Radiative corrections to the neutral-current Drell-Yan process in the Standard Model and its minimal supersymmetric extension}}, \href{https://doi.org/10.1007/JHEP01(2010)060}{\emph{JHEP} {\bfseries 01} (2010) 060}, [\href{https://arxiv.org/abs/0911.2329}{{\ttfamily 0911.2329}}].

\bibitem{Bernaciak:2012hj}
C.~Bernaciak and D.~Wackeroth, \emph{{Combining NLO QCD and Electroweak Radiative Corrections to W boson Production at Hadron Colliders in the POWHEG Framework}}, \href{https://doi.org/10.1103/PhysRevD.85.093003}{\emph{Phys. Rev. D} {\bfseries 85} (2012) 093003}, [\href{https://arxiv.org/abs/1201.4804}{{\ttfamily 1201.4804}}].

\bibitem{Barze:2012tt}
L.~Barze, G.~Montagna, P.~Nason, O.~Nicrosini and F.~Piccinini, \emph{{Implementation of electroweak corrections in the POWHEG BOX: single W production}}, \href{https://doi.org/10.1007/JHEP04(2012)037}{\emph{JHEP} {\bfseries 04} (2012) 037}, [\href{https://arxiv.org/abs/1202.0465}{{\ttfamily 1202.0465}}].

\bibitem{Barze:2013fru}
L.~Barze, G.~Montagna, P.~Nason, O.~Nicrosini, F.~Piccinini and A.~Vicini, \emph{{Neutral current Drell-Yan with combined QCD and electroweak corrections in the POWHEG BOX}}, \href{https://doi.org/10.1140/epjc/s10052-013-2474-y}{\emph{Eur. Phys. J. C} {\bfseries 73} (2013) 2474}, [\href{https://arxiv.org/abs/1302.4606}{{\ttfamily 1302.4606}}].

\bibitem{Frederix:2018nkq}
R.~Frederix, S.~Frixione, V.~Hirschi, D.~Pagani, H.~S. Shao and M.~Zaro, \emph{{The automation of next-to-leading order electroweak calculations}}, \href{https://doi.org/10.1007/JHEP07(2018)185}{\emph{JHEP} {\bfseries 07} (2018) 185}, [\href{https://arxiv.org/abs/1804.10017}{{\ttfamily 1804.10017}}].

\bibitem{Chiesa:2024qzd}
M.~Chiesa, C.~L. Del~Pio and F.~Piccinini, \emph{{On electroweak corrections to neutral current Drell-Yan with the POWHEG BOX}},  \href{https://arxiv.org/abs/2402.14659}{{\ttfamily 2402.14659}}.

\bibitem{deFlorian:2018wcj}
D.~de~Florian, M.~Der and I.~Fabre, \emph{{QCD$\oplus$QED NNLO corrections to Drell Yan production}}, \href{https://doi.org/10.1103/PhysRevD.98.094008}{\emph{Phys. Rev. D} {\bfseries 98} (2018) 094008}, [\href{https://arxiv.org/abs/1805.12214}{{\ttfamily 1805.12214}}].

\bibitem{Delto:2019ewv}
M.~Delto, M.~Jaquier, K.~Melnikov and R.~R\"ontsch, \emph{{Mixed QCD$\otimes$QED corrections to on-shell $Z$ boson production at the LHC}}, \href{https://doi.org/10.1007/JHEP01(2020)043}{\emph{JHEP} {\bfseries 01} (2020) 043}, [\href{https://arxiv.org/abs/1909.08428}{{\ttfamily 1909.08428}}].

\bibitem{Cieri:2020ikq}
L.~Cieri, D.~de~Florian, M.~Der and J.~Mazzitelli, \emph{{Mixed QCD\ensuremath{\otimes}QED corrections to exclusive Drell Yan production using the q$_{T}$ -subtraction method}}, \href{https://doi.org/10.1007/JHEP09(2020)155}{\emph{JHEP} {\bfseries 09} (2020) 155}, [\href{https://arxiv.org/abs/2005.01315}{{\ttfamily 2005.01315}}].

\bibitem{Bonciani:2016wya}
R.~Bonciani, F.~Buccioni, R.~Mondini and A.~Vicini, \emph{{Double-real corrections at $\mathcal{O}(\alpha \alpha_s)$ to single gauge boson production}}, \href{https://doi.org/10.1140/epjc/s10052-017-4728-6}{\emph{Eur. Phys. J. C} {\bfseries 77} (2017) 187}, [\href{https://arxiv.org/abs/1611.00645}{{\ttfamily 1611.00645}}].

\bibitem{Bonciani:2019nuy}
R.~Bonciani, F.~Buccioni, N.~Rana, I.~Triscari and A.~Vicini, \emph{{NNLO QCD$\times$EW corrections to Z production in the $q\bar{q}$ channel}}, \href{https://doi.org/10.1103/PhysRevD.101.031301}{\emph{Phys. Rev. D} {\bfseries 101} (2020) 031301}, [\href{https://arxiv.org/abs/1911.06200}{{\ttfamily 1911.06200}}].

\bibitem{Bonciani:2020tvf}
R.~Bonciani, F.~Buccioni, N.~Rana and A.~Vicini, \emph{{NNLO QCD$\times$EW corrections to on-shell $Z$ production}}, \href{https://doi.org/10.1103/PhysRevLett.125.232004}{\emph{Phys. Rev. Lett.} {\bfseries 125} (2020) 232004}, [\href{https://arxiv.org/abs/2007.06518}{{\ttfamily 2007.06518}}].

\bibitem{Bonciani:2021iis}
R.~Bonciani, F.~Buccioni, N.~Rana and A.~Vicini, \emph{{On-shell Z boson production at hadron colliders through \ensuremath{\cal O}(\ensuremath{\alpha}\ensuremath{\alpha}$_{s}$)}}, \href{https://doi.org/10.1007/JHEP02(2022)095}{\emph{JHEP} {\bfseries 02} (2022) 095}, [\href{https://arxiv.org/abs/2111.12694}{{\ttfamily 2111.12694}}].

\bibitem{Buccioni:2020cfi}
F.~Buccioni, F.~Caola, M.~Delto, M.~Jaquier, K.~Melnikov and R.~R\"ontsch, \emph{{Mixed QCD-electroweak corrections to on-shell Z production at the LHC}}, \href{https://doi.org/10.1016/j.physletb.2020.135969}{\emph{Phys. Lett. B} {\bfseries 811} (2020) 135969}, [\href{https://arxiv.org/abs/2005.10221}{{\ttfamily 2005.10221}}].

\bibitem{Behring:2020cqi}
A.~Behring, F.~Buccioni, F.~Caola, M.~Delto, M.~Jaquier, K.~Melnikov and R.~R\"ontsch, \emph{{Mixed QCD-electroweak corrections to $W$-boson production in hadron collisions}}, \href{https://doi.org/10.1103/PhysRevD.103.013008}{\emph{Phys. Rev. D} {\bfseries 103} (2021) 013008}, [\href{https://arxiv.org/abs/2009.10386}{{\ttfamily 2009.10386}}].

\bibitem{Dittmaier:2014qza}
S.~Dittmaier, A.~Huss and C.~Schwinn, \emph{{Mixed QCD-electroweak $\mathcal{O}(\alpha_s\alpha)$ corrections to Drell-Yan processes in the resonance region: pole approximation and non-factorizable corrections}}, \href{https://doi.org/10.1016/j.nuclphysb.2014.05.027}{\emph{Nucl. Phys. B} {\bfseries 885} (2014) 318--372}, [\href{https://arxiv.org/abs/1403.3216}{{\ttfamily 1403.3216}}].

\bibitem{Dittmaier:2015rxo}
S.~Dittmaier, A.~Huss and C.~Schwinn, \emph{{Dominant mixed QCD-electroweak O($\alpha$$_s$$\alpha$) corrections to Drell\textendash{}Yan processes in the resonance region}}, \href{https://doi.org/10.1016/j.nuclphysb.2016.01.006}{\emph{Nucl. Phys. B} {\bfseries 904} (2016) 216--252}, [\href{https://arxiv.org/abs/1511.08016}{{\ttfamily 1511.08016}}].

\bibitem{Dittmaier:2020vra}
S.~Dittmaier, T.~Schmidt and J.~Schwarz, \emph{{Mixed NNLO QCD$\times$electroweak corrections of $\mathcal{O}(N_f \alpha_s \alpha)$ to single-W/Z production at the LHC}}, \href{https://doi.org/10.1007/JHEP12(2020)201}{\emph{JHEP} {\bfseries 12} (2020) 201}, [\href{https://arxiv.org/abs/2009.02229}{{\ttfamily 2009.02229}}].

\bibitem{Dittmaier:2024row}
S.~Dittmaier, A.~Huss and J.~Schwarz, \emph{{Mixed NNLO QCD $\times$ electroweak corrections to single-Z production in pole approximation: differential distributions and forward-backward asymmetry}},  \href{https://arxiv.org/abs/2401.15682}{{\ttfamily 2401.15682}}.

\bibitem{Bonciani:2021zzf}
R.~Bonciani, L.~Buonocore, M.~Grazzini, S.~Kallweit, N.~Rana, F.~Tramontano and A.~Vicini, \emph{{Mixed Strong-Electroweak Corrections to the Drell-Yan Process}}, \href{https://doi.org/10.1103/PhysRevLett.128.012002}{\emph{Phys. Rev. Lett.} {\bfseries 128} (2022) 012002}, [\href{https://arxiv.org/abs/2106.11953}{{\ttfamily 2106.11953}}].

\bibitem{Buccioni:2022kgy}
F.~Buccioni, F.~Caola, H.~A. Chawdhry, F.~Devoto, M.~Heller, A.~von Manteuffel, K.~Melnikov, R.~R\"ontsch and C.~Signorile-Signorile, \emph{{Mixed QCD-electroweak corrections to dilepton production at the LHC in the high invariant mass region}}, \href{https://doi.org/10.1007/JHEP06(2022)022}{\emph{JHEP} {\bfseries 06} (2022) 022}, [\href{https://arxiv.org/abs/2203.11237}{{\ttfamily 2203.11237}}].

\bibitem{Armadillo:2022bgm}
T.~Armadillo, R.~Bonciani, S.~Devoto, N.~Rana and A.~Vicini, \emph{{Two-loop mixed QCD-EW corrections to neutral current Drell-Yan}}, \href{https://doi.org/10.1007/JHEP05(2022)072}{\emph{JHEP} {\bfseries 05} (2022) 072}, [\href{https://arxiv.org/abs/2201.01754}{{\ttfamily 2201.01754}}].

\bibitem{Bonciani:2016ypc}
R.~Bonciani, S.~Di~Vita, P.~Mastrolia and U.~Schubert, \emph{{Two-Loop Master Integrals for the mixed EW-QCD virtual corrections to Drell-Yan scattering}}, \href{https://doi.org/10.1007/JHEP09(2016)091}{\emph{JHEP} {\bfseries 09} (2016) 091}, [\href{https://arxiv.org/abs/1604.08581}{{\ttfamily 1604.08581}}].

\bibitem{Moriello:2019yhu}
F.~Moriello, \emph{{Generalised power series expansions for the elliptic planar families of Higgs + jet production at two loops}}, \href{https://doi.org/10.1007/JHEP01(2020)150}{\emph{JHEP} {\bfseries 01} (2020) 150}, [\href{https://arxiv.org/abs/1907.13234}{{\ttfamily 1907.13234}}].

\bibitem{Hidding:2020ytt}
M.~Hidding, \emph{{DiffExp, a Mathematica package for computing Feynman integrals in terms of one-dimensional series expansions}}, \href{https://doi.org/10.1016/j.cpc.2021.108125}{\emph{Comput. Phys. Commun.} {\bfseries 269} (2021) 108125}, [\href{https://arxiv.org/abs/2006.05510}{{\ttfamily 2006.05510}}].

\bibitem{Armadillo:2022ugh}
T.~Armadillo, R.~Bonciani, S.~Devoto, N.~Rana and A.~Vicini, \emph{{Evaluation of Feynman integrals with arbitrary complex masses via series expansions}}, \href{https://doi.org/10.1016/j.cpc.2022.108545}{\emph{Comput. Phys. Commun.} {\bfseries 282} (2023) 108545}, [\href{https://arxiv.org/abs/2205.03345}{{\ttfamily 2205.03345}}].

\bibitem{Buonocore:2021tke}
L.~Buonocore, S.~Kallweit, L.~Rottoli and M.~Wiesemann, \emph{{Linear power corrections for two-body kinematics in the $q_T$ subtraction formalism}},  \href{https://arxiv.org/abs/2111.13661}{{\ttfamily 2111.13661}}.

\bibitem{Camarda:2021jsw}
S.~Camarda, L.~Cieri and G.~Ferrera, \emph{{Fiducial perturbative power corrections within the q$_{\bf T}$ subtraction formalism}},  \href{https://arxiv.org/abs/2111.14509}{{\ttfamily 2111.14509}}.

\bibitem{Grazzini:2017mhc}
M.~Grazzini, S.~Kallweit and M.~Wiesemann, \emph{{Fully differential NNLO computations with MATRIX}}, \href{https://doi.org/10.1140/epjc/s10052-018-5771-7}{\emph{Eur. Phys. J. C} {\bfseries 78} (2018) 537}, [\href{https://arxiv.org/abs/1711.06631}{{\ttfamily 1711.06631}}].

\bibitem{Heller:2020owb}
M.~Heller, A.~von Manteuffel, R.~M. Schabinger and H.~Spiesberger, \emph{{Mixed EW-QCD two-loop amplitudes for $q\bar{q} \to \ell^+\ell^-$ and $\gamma_5$ scheme independence of multi-loop corrections}}, \href{https://doi.org/10.1007/JHEP05(2021)213}{\emph{JHEP} {\bfseries 05} (2021) 213}, [\href{https://arxiv.org/abs/2012.05918}{{\ttfamily 2012.05918}}].

\bibitem{Heller:2019gkq}
M.~Heller, A.~von Manteuffel and R.~M. Schabinger, \emph{{Multiple polylogarithms with algebraic arguments and the two-loop EW-QCD Drell-Yan master integrals}}, \href{https://doi.org/10.1103/PhysRevD.102.016025}{\emph{Phys. Rev. D} {\bfseries 102} (2020) 016025}, [\href{https://arxiv.org/abs/1907.00491}{{\ttfamily 1907.00491}}].

\bibitem{Hasan:2020vwn}
S.~M. Hasan and U.~Schubert, \emph{{Master Integrals for the mixed QCD-QED corrections to the Drell-Yan production of a massive lepton pair}}, \href{https://doi.org/10.1007/JHEP11(2020)107}{\emph{JHEP} {\bfseries 11} (2020) 107}, [\href{https://arxiv.org/abs/2004.14908}{{\ttfamily 2004.14908}}].

\bibitem{Buonocore:2021rxx}
L.~Buonocore, M.~Grazzini, S.~Kallweit, C.~Savoini and F.~Tramontano, \emph{{Mixed QCD-EW corrections to $\boldsymbol{pp\!\to\!\ell\nu_\ell\!+\!X}$ at the LHC}}, \href{https://doi.org/10.1103/PhysRevD.103.114012}{\emph{Phys. Rev. D} {\bfseries 103} (2021) 114012}, [\href{https://arxiv.org/abs/2102.12539}{{\ttfamily 2102.12539}}].

\bibitem{Tkachov:1981wb}
F.~V. Tkachov, \emph{{A Theorem on Analytical Calculability of Four Loop Renormalization Group Functions}}, \href{https://doi.org/10.1016/0370-2693(81)90288-4}{\emph{Phys. Lett. B} {\bfseries 100} (1981) 65--68}.

\bibitem{Chetyrkin:1981qh}
K.~G. Chetyrkin and F.~V. Tkachov, \emph{{Integration by Parts: The Algorithm to Calculate beta Functions in 4 Loops}}, \href{https://doi.org/10.1016/0550-3213(81)90199-1}{\emph{Nucl. Phys. B} {\bfseries 192} (1981) 159--204}.

\bibitem{Gehrmann:1999as}
T.~Gehrmann and E.~Remiddi, \emph{{Differential equations for two loop four point functions}}, \href{https://doi.org/10.1016/S0550-3213(00)00223-6}{\emph{Nucl. Phys.} {\bfseries B580} (2000) 485--518}, [\href{https://arxiv.org/abs/hep-ph/9912329}{{\ttfamily hep-ph/9912329}}].

\bibitem{Degrassi:2003rw}
G.~Degrassi and A.~Vicini, \emph{{Two loop renormalization of the electric charge in the standard model}}, \href{https://doi.org/10.1103/PhysRevD.69.073007}{\emph{Phys. Rev.} {\bfseries D69} (2004) 073007}, [\href{https://arxiv.org/abs/hep-ph/0307122}{{\ttfamily hep-ph/0307122}}].

\bibitem{Denner:2005fg}
A.~Denner, S.~Dittmaier, M.~Roth and L.~H. Wieders, \emph{{Electroweak corrections to charged-current e+ e- ---\ensuremath{>} 4 fermion processes: Technical details and further results}}, \href{https://doi.org/10.1016/j.nuclphysb.2011.09.001}{\emph{Nucl. Phys. B} {\bfseries 724} (2005) 247--294}, [\href{https://arxiv.org/abs/hep-ph/0505042}{{\ttfamily hep-ph/0505042}}].

\bibitem{Sirlin:1980nh}
A.~Sirlin, \emph{{Radiative Corrections in the SU(2)-L x U(1) Theory: A Simple Renormalization Framework}}, \href{https://doi.org/10.1103/PhysRevD.22.971}{\emph{Phys. Rev.} {\bfseries D22} (1980) 971--981}.

\bibitem{Kniehl:1989yc}
B.~A. Kniehl, \emph{{Two Loop Corrections to the Vacuum Polarizations in Perturbative QCD}}, \href{https://doi.org/10.1016/0550-3213(90)90552-O}{\emph{Nucl. Phys. B} {\bfseries 347} (1990) 86--104}.

\bibitem{Djouadi:1993ss}
A.~Djouadi and P.~Gambino, \emph{{Electroweak gauge bosons selfenergies: Complete QCD corrections}}, \href{https://doi.org/10.1103/PhysRevD.49.3499}{\emph{Phys. Rev. D} {\bfseries 49} (1994) 3499--3511}, [\href{https://arxiv.org/abs/hep-ph/9309298}{{\ttfamily hep-ph/9309298}}].

\bibitem{Denner:2019vbn}
A.~Denner and S.~Dittmaier, \emph{{Electroweak Radiative Corrections for Collider Physics}}, \href{https://doi.org/10.1016/j.physrep.2020.04.001}{\emph{Phys. Rept.} {\bfseries 864} (2020) 1--163}, [\href{https://arxiv.org/abs/1912.06823}{{\ttfamily 1912.06823}}].

\bibitem{Denner:1994xt}
A.~Denner, G.~Weiglein and S.~Dittmaier, \emph{{Application of the background field method to the electroweak standard model}}, \href{https://doi.org/10.1016/0550-3213(95)00037-S}{\emph{Nucl. Phys. B} {\bfseries 440} (1995) 95--128}, [\href{https://arxiv.org/abs/hep-ph/9410338}{{\ttfamily hep-ph/9410338}}].

\bibitem{Hahn:2000kx}
T.~Hahn, \emph{{Generating Feynman diagrams and amplitudes with FeynArts 3}}, \href{https://doi.org/10.1016/S0010-4655(01)00290-9}{\emph{Comput. Phys. Commun.} {\bfseries 140} (2001) 418--431}, [\href{https://arxiv.org/abs/hep-ph/0012260}{{\ttfamily hep-ph/0012260}}].

\bibitem{Nogueira:1991ex}
P.~Nogueira, \emph{{Automatic Feynman graph generation}}, \href{https://doi.org/10.1006/jcph.1993.1074}{\emph{J. Comput. Phys.} {\bfseries 105} (1993) 279--289}.

\bibitem{Vermaseren:2000nd}
J.~A.~M. Vermaseren, \emph{{New features of FORM}},  \href{https://arxiv.org/abs/math-ph/0010025}{{\ttfamily math-ph/0010025}}.

\bibitem{Maierhofer:2017gsa}
P.~Maierh\"ofer, J.~Usovitsch and P.~Uwer, \emph{{Kira\textemdash{}A Feynman integral reduction program}}, \href{https://doi.org/10.1016/j.cpc.2018.04.012}{\emph{Comput. Phys. Commun.} {\bfseries 230} (2018) 99--112}, [\href{https://arxiv.org/abs/1705.05610}{{\ttfamily 1705.05610}}].

\bibitem{Lee:2013mka}
R.~N. Lee, \emph{{LiteRed 1.4: a powerful tool for reduction of multiloop integrals}}, \href{https://doi.org/10.1088/1742-6596/523/1/012059}{\emph{J. Phys. Conf. Ser.} {\bfseries 523} (2014) 012059}, [\href{https://arxiv.org/abs/1310.1145}{{\ttfamily 1310.1145}}].

\bibitem{Lee:2012cn}
R.~N. Lee, \emph{{Presenting LiteRed: a tool for the Loop InTEgrals REDuction}},  \href{https://arxiv.org/abs/1212.2685}{{\ttfamily 1212.2685}}.

\bibitem{Frenkel:1976bj}
J.~Frenkel and J.~C. Taylor, \emph{{Exponentiation of Leading Infrared Divergences in Massless Yang-Mills Theories}}, \href{https://doi.org/10.1016/0550-3213(76)90320-5}{\emph{Nucl. Phys. B} {\bfseries 116} (1976) 185--194}.

\bibitem{Liu:2022chg}
X.~Liu and Y.-Q. Ma, \emph{{AMFlow: A Mathematica package for Feynman integrals computation via auxiliary mass flow}}, \href{https://doi.org/10.1016/j.cpc.2022.108565}{\emph{Comput. Phys. Commun.} {\bfseries 283} (2023) 108565}, [\href{https://arxiv.org/abs/2201.11669}{{\ttfamily 2201.11669}}].

\bibitem{Catani:1996jh}
S.~Catani and M.~H. Seymour, \emph{{The Dipole formalism for the calculation of QCD jet cross-sections at next-to-leading order}}, \href{https://doi.org/10.1016/0370-2693(96)00425-X}{\emph{Phys. Lett. B} {\bfseries 378} (1996) 287--301}, [\href{https://arxiv.org/abs/hep-ph/9602277}{{\ttfamily hep-ph/9602277}}].

\bibitem{Catani:1996vz}
S.~Catani and M.~H. Seymour, \emph{{A General algorithm for calculating jet cross-sections in NLO QCD}}, \href{https://doi.org/10.1016/S0550-3213(96)00589-5}{\emph{Nucl. Phys. B} {\bfseries 485} (1997) 291--419}, [\href{https://arxiv.org/abs/hep-ph/9605323}{{\ttfamily hep-ph/9605323}}].

\bibitem{Catani:2002hc}
S.~Catani, S.~Dittmaier, M.~H. Seymour and Z.~Trocsanyi, \emph{{The Dipole formalism for next-to-leading order QCD calculations with massive partons}}, \href{https://doi.org/10.1016/S0550-3213(02)00098-6}{\emph{Nucl. Phys. B} {\bfseries 627} (2002) 189--265}, [\href{https://arxiv.org/abs/hep-ph/0201036}{{\ttfamily hep-ph/0201036}}].

\bibitem{Frixione:1995ms}
S.~Frixione, Z.~Kunszt and A.~Signer, \emph{{Three jet cross-sections to next-to-leading order}}, \href{https://doi.org/10.1016/0550-3213(96)00110-1}{\emph{Nucl. Phys. B} {\bfseries 467} (1996) 399--442}, [\href{https://arxiv.org/abs/hep-ph/9512328}{{\ttfamily hep-ph/9512328}}].

\bibitem{Proceedings:2018jsb}
\emph{{Les Houches 2017: Physics at TeV Colliders Standard Model Working Group Report}}, 3, 2018.

\bibitem{Catani:1998bh}
S.~Catani, \emph{{The Singular behavior of QCD amplitudes at two loop order}}, \href{https://doi.org/10.1016/S0370-2693(98)00332-3}{\emph{Phys. Lett. B} {\bfseries 427} (1998) 161--171}, [\href{https://arxiv.org/abs/hep-ph/9802439}{{\ttfamily hep-ph/9802439}}].

\bibitem{Sterman:2002qn}
G.~F. Sterman and M.~E. Tejeda-Yeomans, \emph{{Multiloop amplitudes and resummation}}, \href{https://doi.org/10.1016/S0370-2693(02)03100-3}{\emph{Phys. Lett. B} {\bfseries 552} (2003) 48--56}, [\href{https://arxiv.org/abs/hep-ph/0210130}{{\ttfamily hep-ph/0210130}}].

\bibitem{Becher:2009cu}
T.~Becher and M.~Neubert, \emph{{Infrared singularities of scattering amplitudes in perturbative QCD}}, \href{https://doi.org/10.1103/PhysRevLett.102.162001}{\emph{Phys. Rev. Lett.} {\bfseries 102} (2009) 162001}, [\href{https://arxiv.org/abs/0901.0722}{{\ttfamily 0901.0722}}].

\bibitem{Gardi:2009qi}
E.~Gardi and L.~Magnea, \emph{{Factorization constraints for soft anomalous dimensions in QCD scattering amplitudes}}, \href{https://doi.org/10.1088/1126-6708/2009/03/079}{\emph{JHEP} {\bfseries 03} (2009) 079}, [\href{https://arxiv.org/abs/0901.1091}{{\ttfamily 0901.1091}}].

\bibitem{Kilgore:2011pa}
W.~B. Kilgore and C.~Sturm, \emph{{Two-Loop Virtual Corrections to Drell-Yan Production at order $\alpha_s \alpha^3$}}, \href{https://doi.org/10.1103/PhysRevD.85.033005}{\emph{Phys. Rev. D} {\bfseries 85} (2012) 033005}, [\href{https://arxiv.org/abs/1107.4798}{{\ttfamily 1107.4798}}].

\bibitem{Kilgore:2013uta}
W.~B. Kilgore, \emph{{The Two-Loop Infrared Structure of Amplitudes with Mixed Gauge Groups}}, \href{https://doi.org/10.1140/epjc/s10052-013-2603-7}{\emph{Eur. Phys. J. C} {\bfseries 73} (2013) 2603}, [\href{https://arxiv.org/abs/1308.1055}{{\ttfamily 1308.1055}}].

\bibitem{Becher:2007cu}
T.~Becher and K.~Melnikov, \emph{{Two-loop QED corrections to Bhabha scattering}}, \href{https://doi.org/10.1088/1126-6708/2007/06/084}{\emph{JHEP} {\bfseries 06} (2007) 084}, [\href{https://arxiv.org/abs/0704.3582}{{\ttfamily 0704.3582}}].

\bibitem{Ahmed:2017gyt}
T.~Ahmed, J.~M. Henn and M.~Steinhauser, \emph{{High energy behaviour of form factors}}, \href{https://doi.org/10.1007/JHEP06(2017)125}{\emph{JHEP} {\bfseries 06} (2017) 125}, [\href{https://arxiv.org/abs/1704.07846}{{\ttfamily 1704.07846}}].

\bibitem{Blumlein:2018tmz}
J.~Bl\"umlein, P.~Marquard and N.~Rana, \emph{{Asymptotic behavior of the heavy quark form factors at higher order}}, \href{https://doi.org/10.1103/PhysRevD.99.016013}{\emph{Phys. Rev. D} {\bfseries 99} (2019) 016013}, [\href{https://arxiv.org/abs/1810.08943}{{\ttfamily 1810.08943}}].

\bibitem{Buonocore:2019puv}
L.~Buonocore, M.~Grazzini and F.~Tramontano, \emph{{The $q_T$ subtraction method: electroweak corrections and power suppressed contributions}}, \href{https://doi.org/10.1140/epjc/s10052-020-7815-z}{\emph{Eur. Phys. J. C} {\bfseries 80} (2020) 254}, [\href{https://arxiv.org/abs/1911.10166}{{\ttfamily 1911.10166}}].

\bibitem{Vollinga:2004sn}
J.~Vollinga and S.~Weinzierl, \emph{{Numerical evaluation of multiple polylogarithms}}, \href{https://doi.org/10.1016/j.cpc.2004.12.009}{\emph{Comput. Phys. Commun.} {\bfseries 167} (2005) 177}, [\href{https://arxiv.org/abs/hep-ph/0410259}{{\ttfamily hep-ph/0410259}}].

\bibitem{Ablinger:2010kw}
J.~Ablinger, \emph{{A Computer Algebra Toolbox for Harmonic Sums Related to Particle Physics}}, {\emph{Master's thesis, Linz U.} (2009) }, [\href{https://arxiv.org/abs/1011.1176}{{\ttfamily 1011.1176}}].

\bibitem{Ablinger:2014rba}
J.~Ablinger, \emph{{The package HarmonicSums: Computer Algebra and Analytic aspects of Nested Sums}}, \href{https://doi.org/10.22323/1.211.0019}{\emph{PoS} {\bfseries LL2014} (2014) 019}, [\href{https://arxiv.org/abs/1407.6180}{{\ttfamily 1407.6180}}].

\bibitem{Duhr:2019tlz}
C.~Duhr and F.~Dulat, \emph{{PolyLogTools \textemdash{} polylogs for the masses}}, \href{https://doi.org/10.1007/JHEP08(2019)135}{\emph{JHEP} {\bfseries 08} (2019) 135}, [\href{https://arxiv.org/abs/1904.07279}{{\ttfamily 1904.07279}}].

\bibitem{Hahn:1998yk}
T.~Hahn and M.~Perez-Victoria, \emph{{Automatized one loop calculations in four-dimensions and D-dimensions}}, \href{https://doi.org/10.1016/S0010-4655(98)00173-8}{\emph{Comput. Phys. Commun.} {\bfseries 118} (1999) 153--165}, [\href{https://arxiv.org/abs/hep-ph/9807565}{{\ttfamily hep-ph/9807565}}].

\end{thebibliography}\endgroup
\bibliographystyle{JHEP}

\end{document}